\documentclass{article}

\usepackage{arxiv}

\usepackage[utf8]{inputenc} % allow utf-8 input
\usepackage[T1]{fontenc}    % use 8-bit T1 fonts
\usepackage{hyperref}       % hyperlinks
\usepackage{url}            % simple URL typesetting
\usepackage{booktabs}       % professional-quality tables
\usepackage{amsfonts}       % blackboard math symbols
\usepackage{nicefrac}       % compact symbols for 1/2, etc.
\usepackage{microtype}      % microtypography
\usepackage{lipsum}
\usepackage{graphicx}% Include figure files
\usepackage{dcolumn}% Align table columns on decimal point
\usepackage{bm}% bold math
\usepackage{amsmath}% popular packages from the American Mathematical Society
\usepackage{latexsym}%
\usepackage{lineno}
\usepackage{array}
\usepackage{adjustbox}
\usepackage{algpseudocode}
\usepackage{subfigure}
\usepackage{epsfig}
\usepackage{amsthm}
\usepackage{amssymb}
\usepackage{epstopdf}

\usepackage[dvipsnames]{xcolor}
\usepackage{hyperref}
\usepackage{cleveref}

\definecolor{originalblue}{rgb}{0,0,1}
\definecolor{originalgreen}{rgb}{0,1,0}

\newcommand\myshade{85}
\colorlet{mylinkcolor}{Black}
\colorlet{mycitecolor}{originalgreen}
\colorlet{myurlcolor}{originalblue}

\hypersetup{
  linkcolor  = mylinkcolor!\myshade!black,
  citecolor  = mycitecolor!\myshade!black,
  urlcolor   = myurlcolor!\myshade!black,
  colorlinks = true,
}

\title{Feasibility Analysis for the Problem of Active Near Field/Far Field Acoustic Pattern Synthesis in Free Space and Shallow Water Environments}

\author{
  Chaoxian Qi \\
  Department of Electrical and Computer Engineering\\
  University of Houston\\
  Houston, TX 77004 \\
  \texttt{cqi4@uh.edu} \\
 \And
  Neil Jerome A.~Egarguin \\
  Department of Mathematics\\
  University of Houston,Houston, TX 77004\\
  Institute of Mathematical Sciences and Physics\\
  University of the Philippines Los Ba\~nos, College \\
  Los Ba\~nos, Laguna, Philippines\\
  \texttt{naegarguin1@up.edu.ph} \\
\And
  Daniel Onofrei \\
  Department of Mathematics\\
  University of Houston\\
  Houston, TX 77004 \\
  \texttt{dtonofre@central.uh.edu} \\
\And
  Jiefu Chen \thanks{Corresponding email: jchen82@central.uh.edu }\\
  Department of Electrical and Computer Engineering\\
  University of Houston\\
  Houston, TX 77004 \\
  \texttt{jchen84@uh.edu} \\
}

	\def\Bn{{\bf n}}

	\def\Bx{{\bf x}}
	\def\By{{\bf y}}
	\def\Bz{{\bf z}}

	\def\B0{{\bf 0}}
	\def\calK{{\mathcal K}}
	\def \calD{{\mathcal D}}

	\def \ba {\begin{array}}
	\def \ea {\end{array}}

\DeclareMathOperator*{\argmin}{arg\,min}

\begin{document}
\maketitle

\begin{abstract}
In this paper, a detailed sensitivity analysis of the active manipulation scheme for scalar Helmholtz fields proposed in our previous works, in both free space and constant-depth homogeneous ocean environments, is presented. We apply the method of moments (MoM) together with Tikhonov regularization with Morozov discrepancy principle to investigate the effects of problem parameters variations on the accuracy and feasibility of the proposed active field control strategy. We discuss the feasibility of the active scheme (power budget and control accuracy) as a function of the frequency, the distance between the control region and the active source, the mutual distance between the control regions, and the size of the control region. Process error is considered as well to investigate the possibility of an accurate active control in the presence of manufacturing noise. The numerical simulations show the accuracy of the active field control scheme and indicate some challenges and limitations for its physical implementations.
\end{abstract}

% keywords can be removed

%  End of title page for Preprint option --------------------------------- %
\section{\label{sec:1} Introduction}
The active control of acoustic fields has been extensively explored in the past decades and is an emerging research area in modern acoustics. Compared with passive control schemes, active control techniques own several advantages, such as more flexibility, high accuracy, easy generalization, etc. Instead of using the interaction between sound and specific materials to control the acoustic field, active control strategies characterize an active source so that it is capable of approximating given field patterns in prescribed exterior regions. The current literature has significantly addressed the idea of the active control of Helmholtz scalar fields in broad applications. These include, but not limited to, active noise control~\cite{gabbert2014active, wang2018near, george2013advances, pan2016active, sachau2016development,eggler2019activeWM, jiang2018review}, personal sound zones or multizone sound reproduction~\cite{choi2002generation, elliott2010minimally, cheer2013design, betlehem2011constrained, brannmark2013compensation}, active acoustic cloaking~\cite{eggler2019active, vasquez2009active, Vasquez2011, eggler2019activeWM, broadband_vasquez, Miller:2007:PC}, remote sensing~\cite{elliott2015modeling, jung2017combining} and metamaterial design~\cite{chen2007acoustic, cummer2016controlling, Greenleaf:2009:IIP, Kohn:2012:VPC}. Active sound control techniques are becoming increasingly ubiquitous to enhance sound-based systems ~\cite{Egarguin2018, EgarguinWM2020, Platt2018, AC3, newactr5, Cheer2019, Cheer2016, Nelson1992, Onofrei2014}. Forward and inverse problems for sound in underwater environments have been widely studied in the literature (see monographs~\cite{Kuperman2011},~\cite{MarineAcoustics},~\cite{Keller1997} and references therein). Several comprehensive reviews are available in~\cite{kuo1999active, jiang2018review, betlehem2015personal, norris2015acoustic, kadic2013metamaterials}, discussing the technicalities and recent advances in various active control schemes.

Directional far-field control is of great importance as it serves as the key technique in the aforementioned applications. In this context, in~\cite{dong2019physical,zhang2017directional}, the authors investigated the directional acoustic manipulation via multi-phase forehead structure of the porpoises to realize beamforming. In the same paradigm, the works~\cite{poletti2013design, rafaely2011optimal} made use of loudspeaker arrays to achieve a directional source.

The majority of the active control strategies have been focused on the method of pressure matching (PM)~\cite{betlehem2015personal}. The PM approach aims to match the target field pattern in the given region with minimum error. The active control problem is cast as an inverse source problem (ISP).

Unlike in the free space environment, the active sound control in underwater environments is much more complicated as the underwater channel poses serious challenges. We mention here the work~\cite{gussen2016survey} where the authors discuss the paradigm of subsea wireless communication,~\cite{Zora2011} for a national security and defense discussion and the works~\cite{Buck1997, peng2014study} where the authors develop a single-mode excitation with a feedback control algorithm to realize both near and far-field sound control.

A general ocean environment can be modeled as a horizontally stratified waveguide~\cite{makris2001unified, Kuperman2011, Keller1997}. In general it is fairly difficult to find the analytical fundamental solution and thus, in this paper we follow the paradigm proposed in~\cite{MarineAcoustics} and consider a simpler marine environment modeled as a shallow water or a homogeneous finite-depth ocean.  Our sensitivity analysis builds up on the numerical framework developed in~\cite{Platt2018, Onofrei2014, Onofrei-S,egarguin2020active}. We use the associated Green's function to represent the solution to the Helmholtz equation and employ the integral equation (IE) method to formulate the forward propagator. The method of moments (MoM) approach is used to reduce the original integral equation to a discretized linear system. Then, a Tikhonov regularization scheme with the Morozov discrepancy principle is applied to solve the resulting system of equations. In the underwater environment, additional boundary conditions need to be considered. Consequently, the Green's function should be modified in the formulation of the forward propagator. We use the normal mode representation to formulate the Green's function in the homogeneous finite-depth ocean~\cite{Kuperman2011, MarineAcoustics}.

In this paper, we present a detailed sensitivity study for the problem of controlling three-dimensional scalar Helmholtz fields in several prescribed exterior regions while maintaining desired far-field pattern values in given fixed directions. We discuss the feasibility of the active scheme (power budget, control accuracy and process error) with respect to variations in frequency, the distance between the control region and the active source, the mutual distance between the control regions, and the control region size.

The rest of this paper is organized as follows. In Section~\ref{sec:2}, we formally describe the problem and provide relevant theoretical results obtained in~\cite{egarguin2020active}. Section~\ref{sec:3} shows the numerical results and sensitivity analysis in free space. In Section~\ref{sec:4}, the numerical results and sensitivity analysis in the shallow water environment are presented. Finally, we conclude the paper with some remarks in Section~\ref{sec:5}.

%========================================================================================
\section{\label{sec:2} Theory}
\subsection{\label{subsec:2:1} Problem formulation}
In this section, we present a general description of the active manipulation scheme for Helmholtz fields proposed in our previous works. The unified geometric and functional framework has already been discussed in~\cite{Onofrei2014, Platt2018, egarguin2020active}. We shall briefly recall several essential theoretical results and geometric configurations.

\begin{figure}[!b]
\centering
\includegraphics[width=0.7\linewidth]{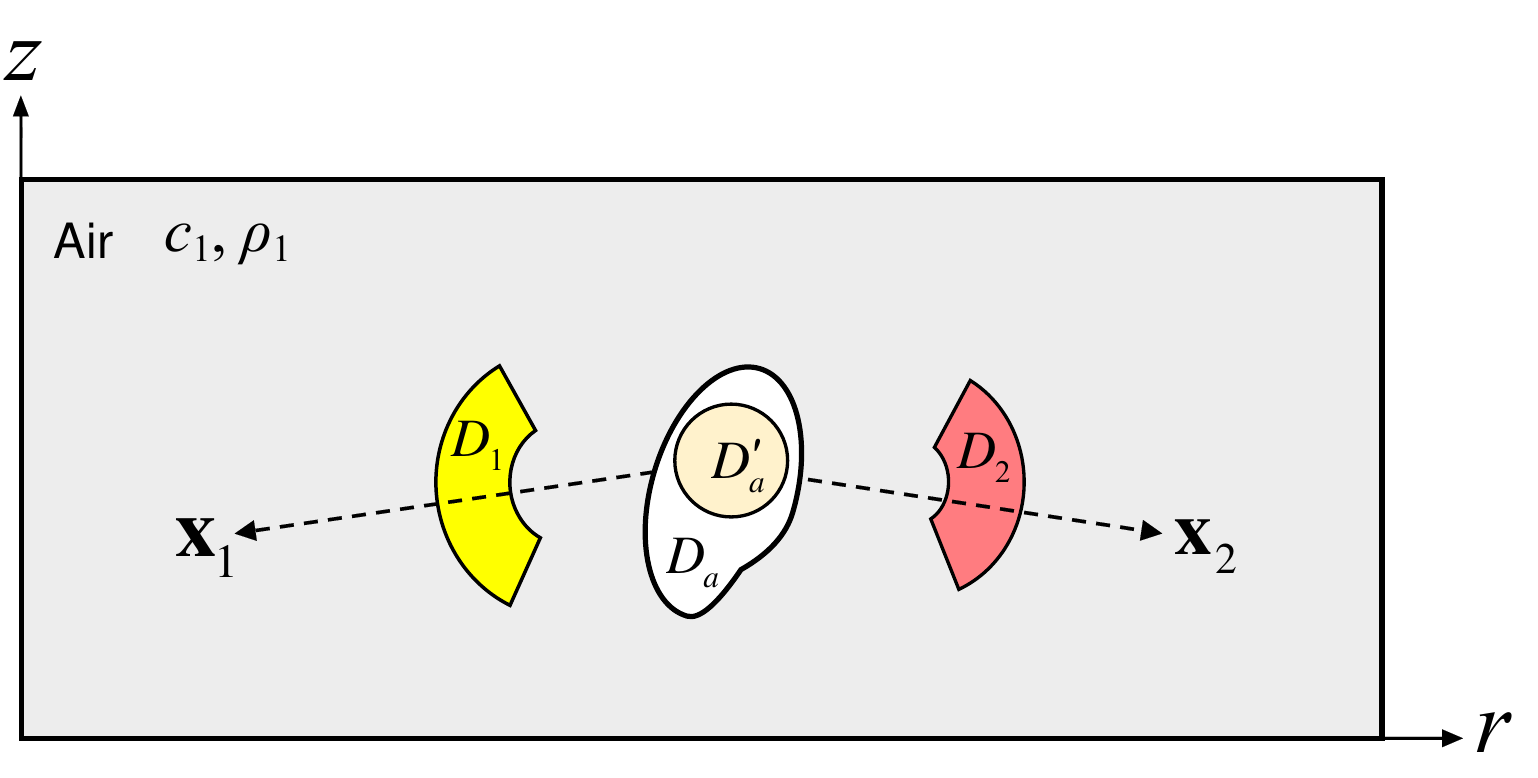}
\caption{Sketch  of the problem geometry showing the near controls $D_1$, $D_2$ and the far field directions $\mathbf{x_1}$ and  $\mathbf{x_2}$ in free space.}
\label{Freespace sketch}
\end{figure}

\begin{figure}[!t]
\centering
\includegraphics[width=0.7\linewidth]{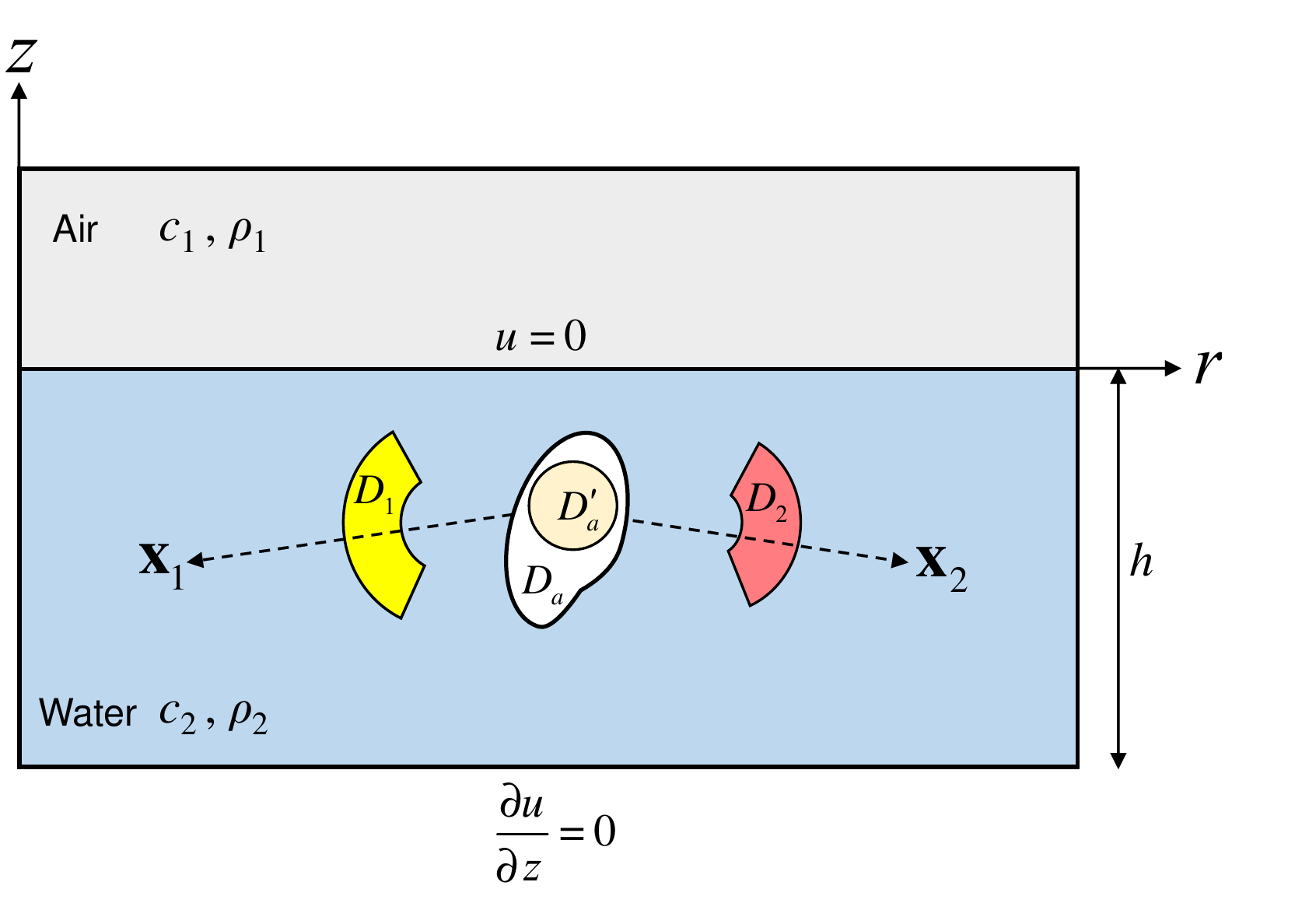}
\caption{Sketch  of the problem geometry showing the near controls $D_1$, $D_2$ and the far field directions $\mathbf{x_1}$ and  $\mathbf{x_2}$ in homogeneous ocean with constant depth.}
\label{Shallow water sketch}
\end{figure}

The problem is to characterize an active source (modeled as surface pressure or surface normal velocity) so that its generated field approximate some prescribed fields in several exterior regions of interest while maintaining desired patterns in several given far field directions. In this paper, the active field manipulation scheme is explored in both free space and homogeneous ocean with a constant depth environments. The geometries of the problem in free space and homogeneous ocean are sketched in Fig.~\ref{Freespace sketch} and Fig.~\ref{Shallow water sketch}, respectively. Although the theoretical discussion in~\cite{Onofrei2014, Platt2018, egarguin2020active} shows that an arbitrary number of source regions, exterior control regions, and far-field directions can be considered in the active scheme, for exemplification we only consider here a single source $D_a$, two control regions $D_1$, $D_2$, and two far-field directions $\mathbf{x_1}$, $\mathbf{x_2}$ for illustrative purposes. A single source $D_a\Subset \mathbb{R}^3$ is modeled as a compact region in both free space and homogeneous ocean. The control regions $D_1$ and $D_2$ are mutually disjoint smooth domains, i.e., $D_1 \cap D_2 = \emptyset$. We assume that the control regions are well-separated from the source region, i.e., $(D_1 \cup D_2) \cap D_a= \emptyset$. Furthermore, we consider two distinct directions $\Bx_1$ and $\Bx_2$ representing the far-field directions of interest.

Mathematically the problem is to find the boundary input on the source, either a Dirichlet input data $p$  (pressure)  or a Neumann input data $v_n$ (normal velocity) such that for any desired field $f = (f_1, f_2)$ on the control regions $D_1$, $D_2$ and prescribed far field pattern values $f_\infty = (f_{\infty,1}, f_{\infty,2})$, the solution $u$ of the following exterior Helmholtz problem:\begin{equation}
\label{Helmholtz problem}
\left\{
\begin{array}{llll}
\nabla^2 u+k^2u=0 \mbox{ in } \mathbb{R}^3 \!\setminus D_a \vspace{0.15cm},\\
\nabla u\cdot \Bn=v_n, (\mbox{ or } u=p)\mbox{ on }\partial D_a,\\
\text{corresponding boundary conditions},\\
\text{suitable radiation condition}
 \end{array}\right.
\end{equation}
satisfies the control constraints
\begin{equation}
\label{Error threshold}
\begin{cases}
\Vert u - f_l\Vert_{C^2(D_l)}\leq \mu \text{ for } l= \overline{1,2} \: , \\
|u_\infty(\hat \Bx_j)-f_{\infty,j}| \leq \mu \text{ for } j= \overline{1,2} \: ,
\end{cases}
 \end{equation}
where $0 < \mu \ll 1$ is the desired control accuracy threshold and $u_\infty$ denotes the the far field pattern of $u$. In~(\ref{Helmholtz problem}) and~(\ref{Error threshold}), $\Bn$ denotes the outward pointing vector normal to $\partial D_a$ and ${\hat{\Bx}}=\frac{\Bx}{|\Bx|}$ is the unit vector along the direction $\Bx$. The subscript $C^2$ denotes the norm in the space of smooth functions with continuous derivatives up to the second order computed as a sum of the $L^2$ norms of all the partial derivatives up to the second order where the $L^2$ norm on the space of square integrable functions on a given domain $D$ is defined as $
\Vert f \Vert^2_{L^2(D)} = \int_{D} |f(\mathbf x)|^2 ~d{\mathbf x}$. The boundary conditions and radiation condition in~(\ref{Helmholtz problem}) are corresponding to the medium (environment) and will be given in Section~\ref{subsec:2:2} and Section~\ref{subsec:2:3}.

In~\cite{Onofrei2014,Platt2018}, it was demonstrated that problem~(\ref{Helmholtz problem}) together with~(\ref{Error threshold}) admits a solution if the wavenumber $k$ is not a resonance. As can be observed in Fig.~\ref{Freespace sketch} and Fig.~\ref{Shallow water sketch}, in order to ease the analysis and integral computations our scheme makes use of a "fictitious source", i.e., an arbitrary sphere $D'_a$ compactly embedded in the actual source region $D_a$. In general, the physical source $D_a$ can have any shape as long as it has a  Lipschitz boundary, compactly includes the fictitious source  $D'_a$ and $(D_1 \cup D_2) \cap D_a= \emptyset$. Meanwhile, our scheme uses slightly larger mutually disjoint regions $W_1$, $W_2$ such that $D_1 \Subset W_1$, $D_2 \Subset W_2$, $W_1 \cap W_2 = \emptyset$ and $(W_1 \cup W_2) \cap D_a= \emptyset$ because, as shown in \cite{Onofrei2014}, $L^2$ control on $\partial W_1, \partial W_2$ imply, via regularity and uniqueness results for the solution of interior Helmholtz, the sooth control stated in \eqref{Error threshold}.  As pointed out in~\cite{Platt2018, egarguin2020active}, within the framework mentioned above, the boundary input data, either normal velocity $v_n$ or pressure $p$ on the surface of the active source can be characterized using a smooth density function $w \in L^2(\partial D'_a)$ such that
\begin{eqnarray}
& v_n(\mathbf x) =&\displaystyle \frac{-i}{\rho c k}\frac{\partial}{\partial\Bn}\int_{\partial D'_{a}}w(\By) \phi(\Bx, \By) dS_\By,
\label{eqnvn}\\
\nonumber \\
& p(\mathbf x) =&\displaystyle \int_{\partial D'_{a}}w(\By) \phi(\Bx,\By) dS_\By,
\label{eqnpb}
\end{eqnarray}
for $\Bx\in\partial D_a$ and where $\rho$ is the density of the surrounding environment, $c$ is the speed of sound in the given medium and $\phi(\Bx, \By)$ is the fundamental solution of the 3D Helmholtz equation.  In general, the solution of problem \eqref{Helmholtz problem} can be represented by a linear combination of a single and a double layer potential illustrated in~\cite{ahrens2010single, colton_kress}. For simplicity of the computations, we only use the single-layer potential operator throughout this paper.

%===========================================================================
\subsection{\label{subsec:2:2} Free space environment}
In this section, the active manipulation of Helmholtz fields in a free-space environment is investigated. In this medium, problem~(\ref{Helmholtz problem}) states
\begin{equation}
\label{Helmholtz problem free space}
\vspace{0.15cm}\left\{\vspace{0.15cm}
\begin{array}{llll}
\nabla^2 u+k^2u=0 \mbox{ in } \mathbb{R}^3 \!\setminus D_a \vspace{0.15cm},\\
\nabla u\cdot \Bn=v_n, (\mbox{ or } u=p)\mbox{ on }\partial D_a,\\
%\text{corresponding boundary conditions}\\
% \text{suitable radiation condition}
\left
<{\hat{\Bx}},\nabla u(\Bx)\right>\! -\!iku(\Bx)\!=\!o\left(\frac{1}{|\Bx|}\!\right)\!,\mbox{ as }|\Bx|\rightarrow\infty.
\end{array}
\right.
\end{equation}
The fundamental solution in this case is given by $\phi(\Bx, \By) = \dfrac{e^{ik|\Bx-\By|}}{4 \pi |\Bx-\By|}$. Then, the control problem is to characterize $p$ or $v_n$ such that $u$ of~(\ref{Helmholtz problem free space}) satisfies \eqref{Error threshold}. In free space, solution $u$ can be written as a propagator operator $\calD$ in terms of the density function $w$. Indeed, one can define the propagator operator $\calK_l$ on the control region $\partial W_l$ as
\begin{equation}
\calK_l w (\Bz_l) = \displaystyle \int_{\partial D'_{a}}w(\By) \phi(\Bz_l,\By) dS_\By,
\end{equation}
where for each $l =\overline{1,2}$, $\Bz_l \in \partial W_l$ and $\By \in \partial D'_a$. Following the derivation in~\cite{colton_kress}, the far field pattern operator $\calK_{\infty,j}$ can be defined as
\begin{equation}
\calK_{\infty,j} w (\hat \Bx_j) = \dfrac{1}{4\pi} \int_{\partial D_a'} w(\By) e^{-i k \mathbf{\hat \Bx_j} \cdot \By} dS_\By,
\end{equation}
where $\hat \Bx_j$, $j =\overline{1,2}$ is the unit vector pointing in the far-field direction of interest. Hence, the overall propagator operator $\calD$ in free space is defined as
\begin{equation}
\begin{split}
    \calD w (\Bz_1, \Bz_2, \hat \Bx_1 , \hat \Bx_2)  = & \big ( \calK_1 w (\Bz_1),  \calK_2 w (\Bz_2),  \calK_{\infty,1} w (\hat \Bx_1), \calK_{\infty,2} w (\hat \Bx_2) \big ).
\end{split}
\label{propagator free}
\end{equation}

%=============================================================================
\subsection{\label{subsec:2:3} Homogeneous ocean environment}
Compared with the free space regime, the active control scheme in the homogeneous ocean with a constant depth is much more complicated. As shown in Fig.~\ref{Shallow water sketch}, two additional boundary conditions need to be considered. More explicitly, problem~(\ref{Helmholtz problem}) now reads
\begin{equation}
\vspace{0.15cm}\left\{\vspace{0.15cm}\begin{array}{llll}
\nabla^2 u+k^2u=0 \mbox{ in } \mathbb{R}^3 \!\setminus D_a \vspace{0.15cm},\\
\nabla u\cdot \Bn=v_n, (\mbox{ or } u=p)\mbox{ on }\partial D_a,\\
u=0 \text{ at the ocean surface } z=0 , \\
\dfrac{\partial u}{\partial z} = 0 \text{ at the ocean floor }z=h ,\\
\displaystyle \lim_{r \to \infty} r^{1/2} \left (\dfrac{\partial u_p}{\partial r} -ika_pu_p\right ) = 0, \text{ for } \theta \in [0, 2 \pi).
\end{array}\right.
\label{Helmholtz problem in homogeneous ocean}
\end{equation}
where following the framework proposed in ~\cite{MarineAcoustics}, we employ cylindrical coordinates in our analysis and the functions $u_p$'s represent the normal modes in the representation of $u$. The main control problem is then to characterize $v_n$ or $p$ such that $u$ satisfies \eqref{Error threshold}. We employ the following Green's representation for $u$,
\begin{equation}
\label{sl-ocean}
u(\Bx) = \int_{\partial D'_a}  w(\By) G(\Bx, \By) dS_\By,
\end{equation}
where $w$ is the density function defined on the fictitious surface source $\partial D_a'$ and $G$ is the associated Green's function in the medium. For any observation point $\mathbf x = (r, \theta, z) = (\mbox{\boldmath$\xi$}, z)$ and source point $\mathbf y = (r', \theta', z') = (\mbox{\boldmath$\xi$}', z')$, the Green's function has the following normal mode representation:
\begin{equation}
\label{Green's function in ocean}
G(\mathbf x, \mathbf y) = \dfrac{i}{2h} \sum_{p=0}^{+\infty} \phi_p(z) \phi_p(z') H_0^{(1)}(ka_p|\mbox{\boldmath$\xi$}-\mbox{\boldmath$\xi$}'|),
\end{equation}
where $H_0^{(1)}(x)$ is the zero order Hankel function of the first kind, $\phi_p$ is the $p^{\text{th}}$ modal solution with associated eigenvalue $a_p$~\cite{Keller1997,MarineAcoustics,Kuperman2011}. These eigenvalues are
\begin{equation}
\label{a_n}
a_p = \sqrt{1-\dfrac{(2p+1)^2\pi^2}{4k^2h^2}},
\end{equation}
while the separated modal solutions $\phi_p$ are given by
\begin{equation}
\label{phi_n}
\phi_p(z) = \sin \left [ k \sqrt{1-a_p^2}z\right].
\end{equation}

In the far-field region, the field $u$ has an asymptotic form given by~\cite{MarineAcoustics},
\begin{equation}
u(\Bx) = \sum_{p=0}^N \dfrac{1}{\sqrt{ka_pr}} e^{ika_pr}g_p(\theta, z) + \mathcal O \left (\dfrac{1}{r^{3/2}} \right ), \text{ as } r \to +\infty,
\end{equation}
where $N$ is the number of propagating modes (i.e., the larges integer so that $a_p\in{\mathbb R}$), $g_p$ is given by
\begin{equation}
\begin{split}
    g_p( \theta, z) = & \sqrt{\dfrac{2}{\pi}} \int_{\partial D_a'}  w(\By) \cdot \left ( \sum_{q=0}^{\infty} e^{-i(q+\frac{1}{2})\frac{\pi}{2}} \alpha_{qp}(z, \theta, r', z', \theta')\right ) dS_\By,
\end{split}
\end{equation}
and
\begin{equation}
\begin{split}
    \alpha_{qp}(z, \theta, r', z', \theta') = & \dfrac{i \epsilon_q}{2h} \phi_p(z) \cdot  \left [ \cos(q \theta) \beta_{qp}(\By) + \sin(q \theta) \gamma_{qp}(\By) \right],
\end{split}
\end{equation}
where $\epsilon_0=1$, and $\epsilon_q=2$ for $q\geq 1$ with
\begin{align}
\beta_{qp}(\By) &=J_q(ka_pr')\phi_p(z')\cos (q \theta'), \\
\gamma_{qp}(\By) &= J_q(ka_pr')\phi_p(z')\sin (q \theta'),
\end{align}
where $J_p(x)$ is the Bessel function of the first kind of order $p$. Therefore, the far-field pattern in a given direction $\hat \Bx = (1,\theta, z)$ can be defined as~\cite{MarineAcoustics}
\begin{equation}
\label{far field pattern}
u_\infty(\hat \Bx) = \sum_{p=0}^N g_p( \theta, z).
\end{equation}

Similar to the free space regime, we define a propagator operator  $\calD$ that calculates the generated field on the control regions and the far-field pattern in the given directions. For each $l =\overline{1,2}$ and $j =\overline{1,2}$, define
\begin{equation}
\calK_l w (\Bz_l) = \displaystyle \int_{\partial D'_{a}}w(\By) G(\Bz_l,\By) dS_\By \text{ and}
\end{equation}
\begin{equation}
\calK_{\infty,j} w (\hat \Bx_j) =  \sum_{p=0}^N g_p( \theta_j, z_j).
\end{equation}
The overall propagator operator $\calD$ is then given by
\begin{equation}
\begin{split}
    \calD w (\Bz_1, \Bz_2, \hat \Bx_1 , \hat \Bx_2)  = & \big ( \calK_1 w (\Bz_1),  \calK_2 w (\Bz_2), \calK_{\infty,1} w (\hat \Bx_1), \calK_{\infty,2} w (\hat \Bx_2) \big ).
\end{split}
\label{propagator ocean}
\end{equation}
%=============================================================================

\subsection{\label{subsec:2:4} Optimization scheme}
In Section~\ref{subsec:2:2} and Section~\ref{subsec:2:3}, we already defined the propagator operator $\calD$ that evaluates the field in the exterior control regions and the far-field pattern values.  The problem~(\ref{Helmholtz problem}), \eqref{Error threshold}, formulated in the respective context of free space model ~(\ref{Helmholtz problem free space}) or constant-depth homogeneous ocean model~(\ref{Helmholtz problem in homogeneous ocean}) can be summarized as
\begin{equation}
    \calD w \approx f.
    \label{general linear system}
\end{equation}
Following the approach in~\cite{EgarguinWM2020} the density function $w$ in~(\ref{general linear system}) is determined using the method of moments (MoM) by discretizing the control regions into a discrete mesh of collocation points and $w$ being expressed as a linear combination (with unknown coefficients) of local basis functions spanning the space of square integrable functions on $\partial D_{a'}$.  Thus, the integral form in~(\ref{general linear system}) is reduced to a linear system,
\begin{equation}
    A w_d = b,
    \label{MoM linear system}
\end{equation}
where $w_d$ represents the vector of unknown coefficients in the local basis representation of $w$, $A$ represents the matrix of moments computed from the propagator $\cal{D}$ and $b$ is the vector of values of $f$ in the mesh of evaluation points distributed within the control regions together with the prescribed two far field directions.  The matrix $A$ is not invertible in most cases, thus the linear system~(\ref{MoM linear system}) is solved using a regularization routine to minimize the sum of squared residuals. Following the strategy in~\cite{Onofrei2014,Onofrei-S,Egarguin2018,Platt2018}, the unknown coefficients in $w_d$ are obtained by using Tikhonov regularization, and can be compactly written as
\begin{equation}
    \hat w_d = \argmin_{w_d \in \partial D'_a} \Vert Aw_d - b \Vert_{L^2(\partial W_l)}^2 + \alpha \Vert w_d  \Vert_{L^2(\partial D'_a)}^2,
    \label{TRLS}
\end{equation}
where $\alpha$ is the regularization parameter representing the penalty weight for the power required by the solution. The optimal $\alpha$ is determined by the Morozov discrepancy principle~\cite{bonesky2008morozov, ColtonKress2013}. The discrete unknown coefficients $w_d$ are evaluated as the Tikhonov solution,
\begin{equation}
    w_d = (\alpha I + A^*A)^{-1}A^*b,
    \label{TRLS solution}
\end{equation}
where $I$ is the identity matrix and $A^*$ is the complex conjugate transpose of $A$.

To estimate the power on the actual source $D_a$, the averaged radiated power $P_{ave}$ and the stored energy $P_{stor}$ are computed as
\begin{align}
    \label{Average power}
    P_{ave} &= \frac{1}{2} \int_{\partial B_R} \text{Re} [ u^* (\nabla u \cdot \Bn) ] ~dS \text{ and} \\
    P_{stor} &= \frac{1}{2} \int_{\partial B_R} \text{Im} [ u^* (\nabla u \cdot \Bn) ] ~dS,
    \label{Stored energy}
\end{align}
where $u^*$ denotes the complex conjugate and $B_R$ is some sphere of radius $R$ containing the actual source $D_a$. In our analyses, the calculated power is expressed in dB relative to a reference level of $10^{-12}$ $W$. Finally, the sought boundary input: either normal velocity $v_n$ or pressure $p$ on the actual source is obtained from~(\ref{eqnvn}) and~(\ref{eqnpb}), respectively.
%=============================================================================
\section{\label{sec:3} Numerical results in the free space}
\begin{figure}[!b]
\centering
    \subfigure[]{
        \includegraphics[width=0.35\linewidth]{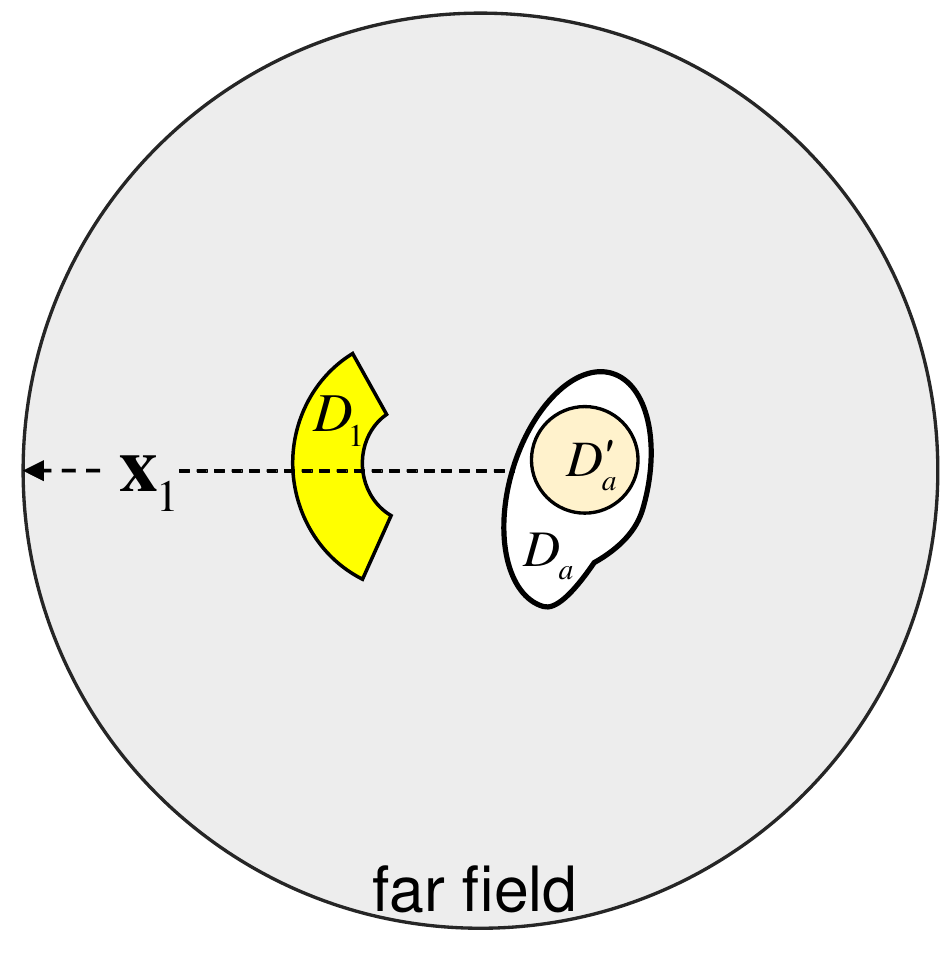}}
  \subfigure[]{
        \includegraphics[width=0.35\linewidth]{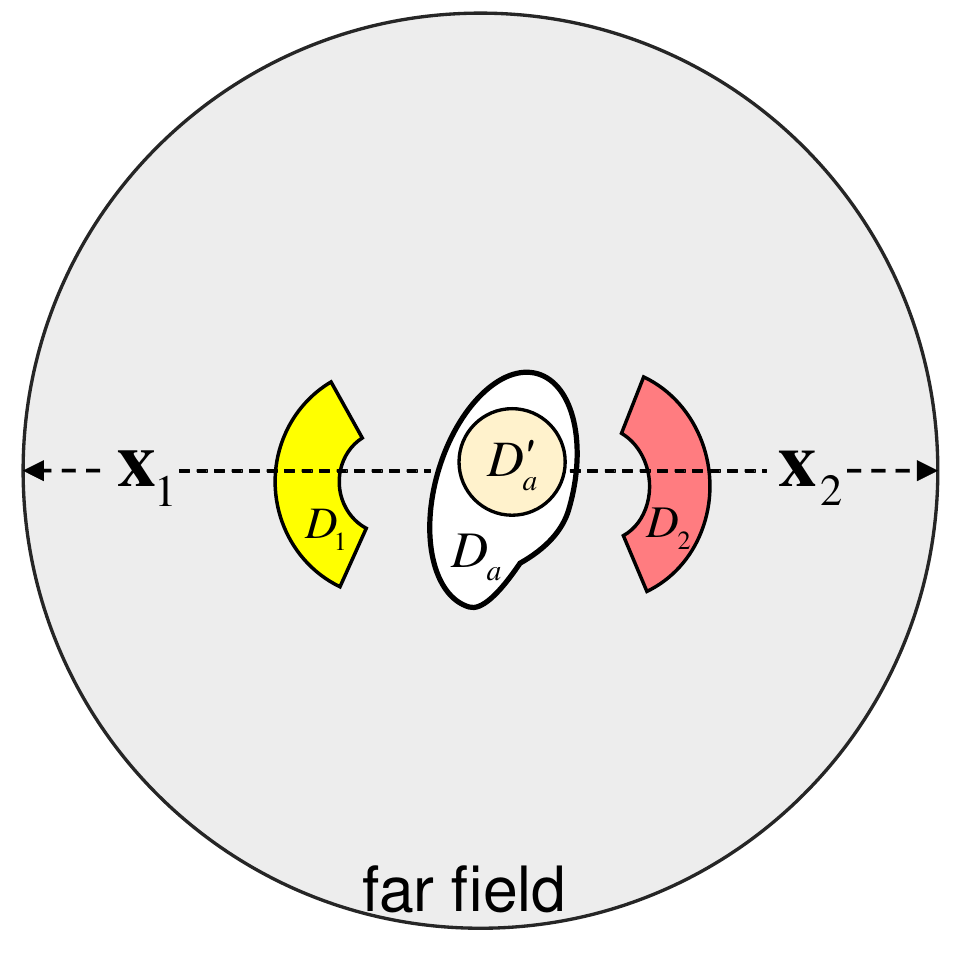}}
\caption{Sketch  of the top view of the problem geometry showing the near control(s) and the far field direction(s). (a) One near control and one far field direction. (b) Two near controls and two far field directions.}
\label{Geometry Top view}
\end{figure}

In this section, we present several relevant numerical simulations to support the above mentioned theoretical framework. We start from a simplified geometric configuration as shown in Fig.~\ref{Geometry Top view}(a) with one near control region $D_1$ and one far-field direction ${\Bx_1}$ which is exactly behind the near control. Then, we extend our numerical study into a multiple-region regime with two near field control regions $D_1$ and $D_2$ and two far field directions ${\Bx_1}$ behind $D_1$ and ${\Bx_2}$ behind $D_2$ as sketched in Fig.~\ref{Geometry Top view}(b). The active source and control regions are in free space (with medium parameters $c$ = 343 m/s and $\rho$ = 1.225 kg/m$^3$). Throughout this section, the fictitious source region is the sphere of radius 0.2 m centered at the origin. In general, the actual source $D_a$ can be arbitrarily shaped as long as it is Lipschitz and compactly embeds $D'_a$. In our simulations, for exemplification, we assume the actual source to be the sphere of radius 0.22 m centered at the origin. The section starts with Subsection~\ref{subsec:3:1}, and Subsection~\ref{subsec:3:2}  which, in the spirit of~\cite{egarguin2020active}, discuss the performance of our strategy in each of the above mentioned configurations and then continues with Section~\ref{subsec:3:3} where we present a detailed sensitivity analysis for free space.
% ==================================================================================
\subsection{\label{subsec:3:1} A null near control and non-zero far field pattern}
\begin{figure}[!t]
\centering
    \includegraphics[width=0.5\linewidth]{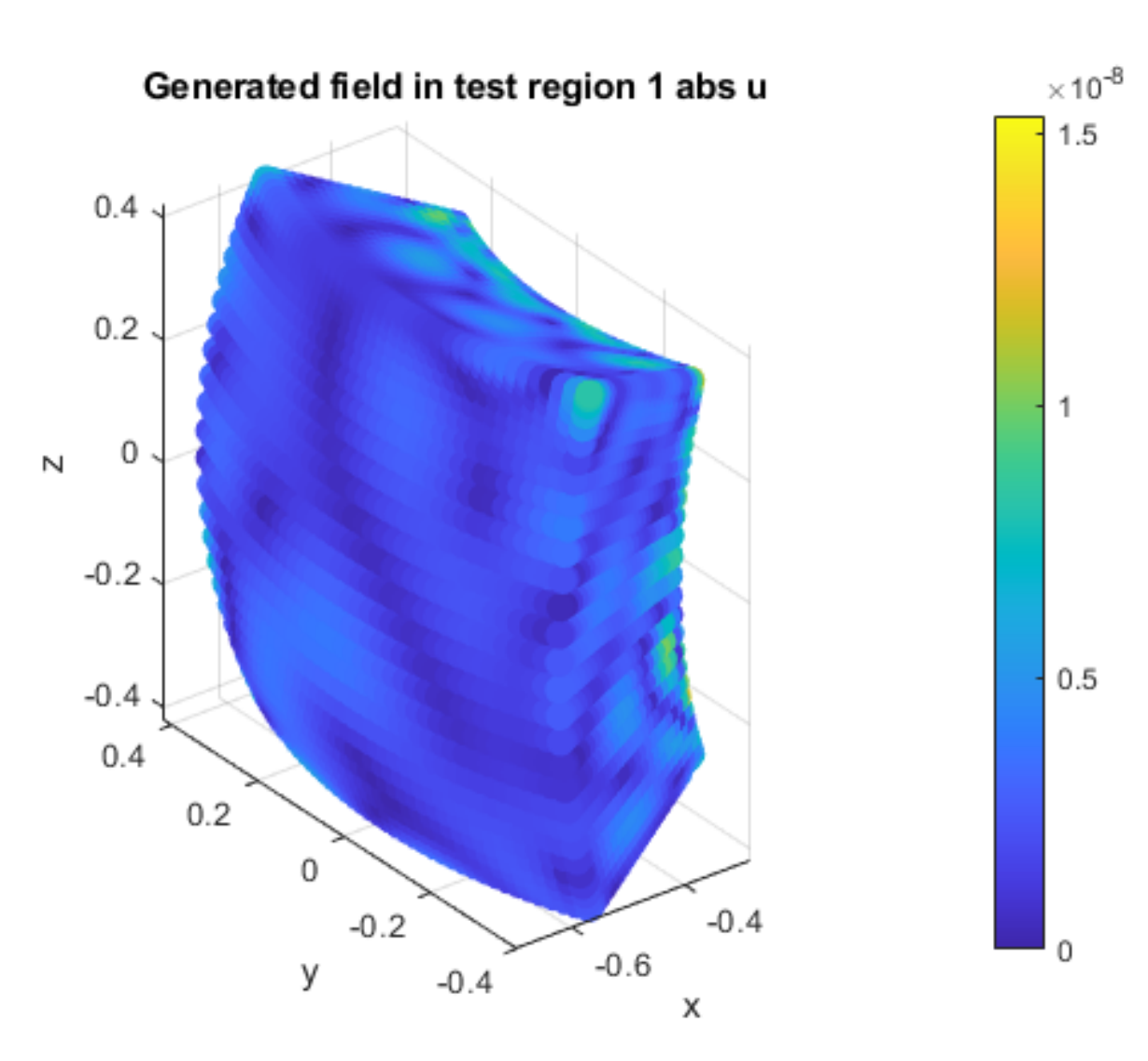}
\caption{Pointwise magnitude of the generated field in the near control approximating a null field.}
\label{Free Space Near Control}
\end{figure}

\begin{figure}[!t]
\centering
    \subfigure[]{
        \includegraphics[width=0.4\linewidth]{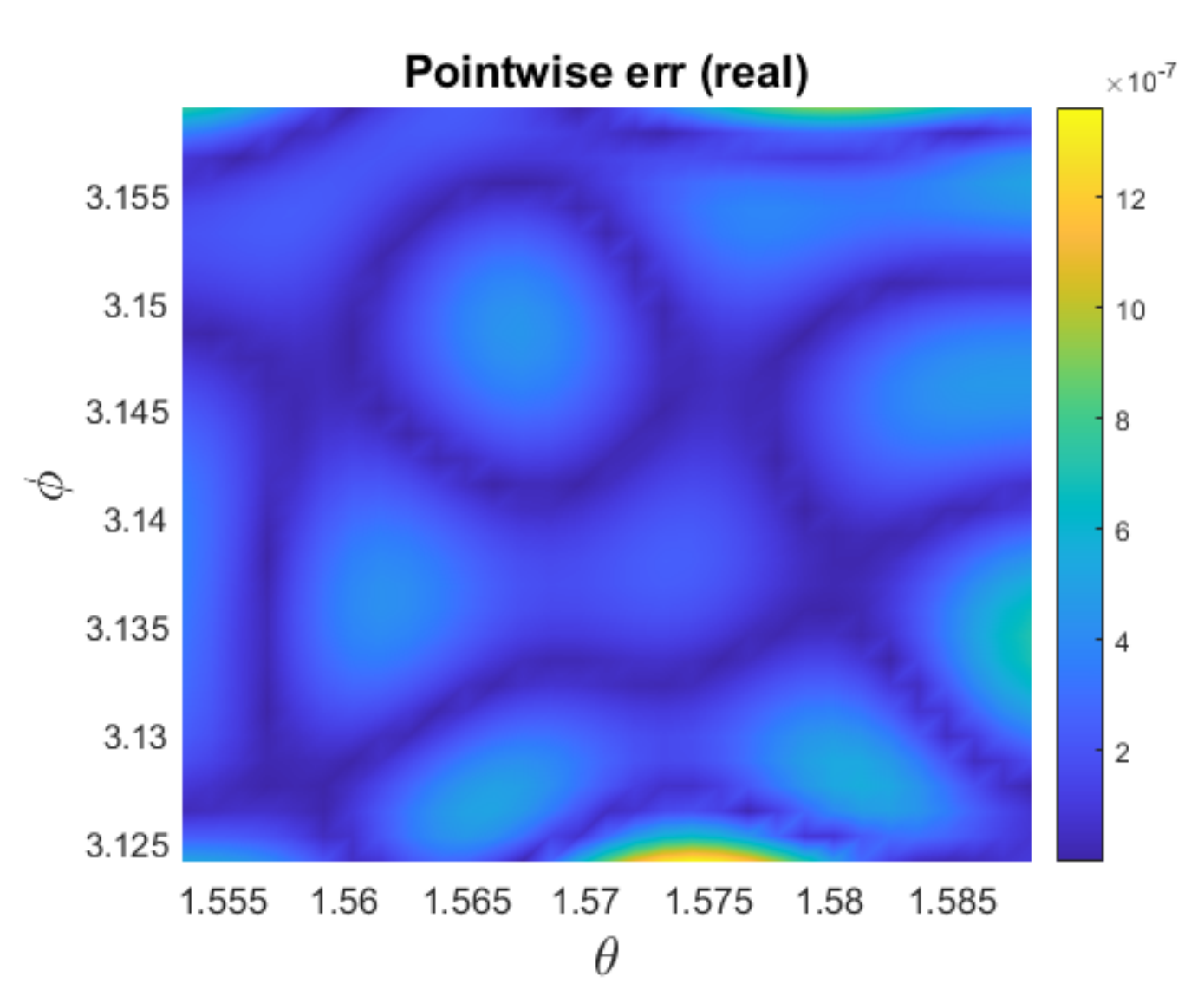}}
    \subfigure[]{
        \includegraphics[width=0.4\linewidth]{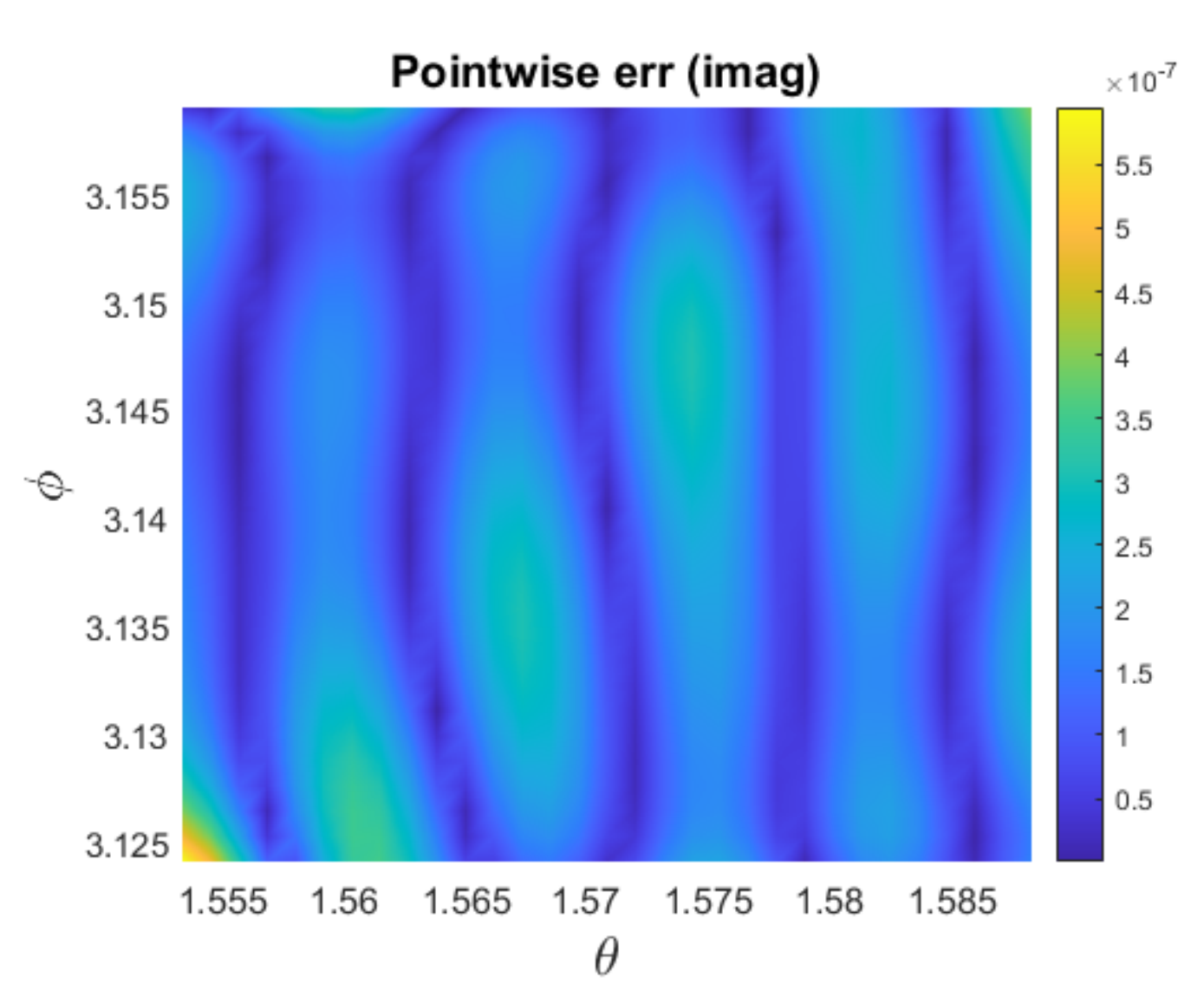}}
\caption{Pointwise relative error in a patch around the far field direction: (a) Real part, (b) Imaginary part.}
\label{Free Space Far Control}
\end{figure}

\begin{figure}[!b]
\centering
\includegraphics[width=0.5\linewidth]{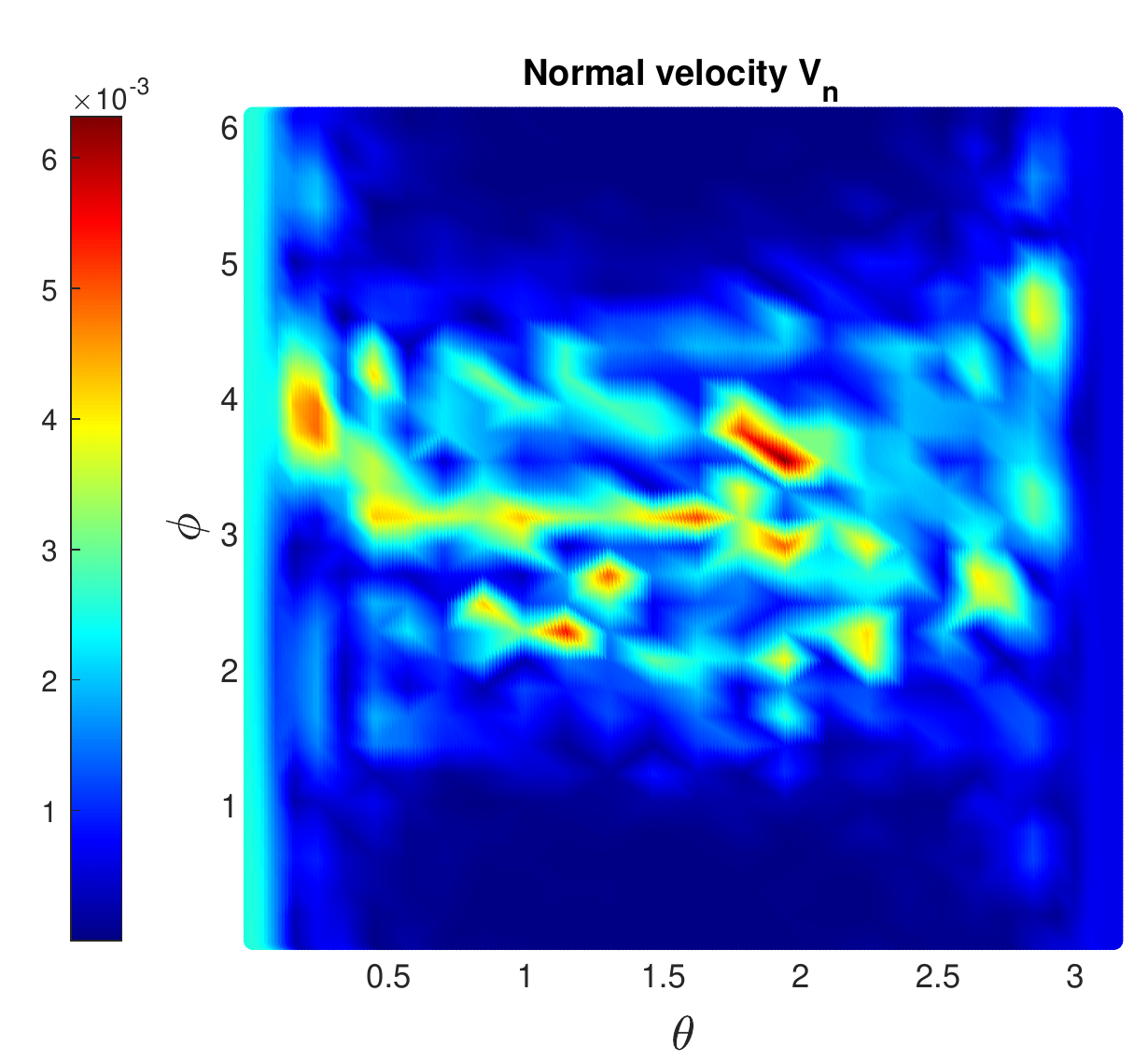}
\caption{Boundary input $v_n$ on the actual source $\partial D_a$.}
\label{Surface Normal velocity}
\end{figure}

In this subsection we show the performance of our scheme in creating a null in  $D_1$ and a given pattern $f_{\infty , 1}$ in the prescribed far field direction $\Bx_1$ (see Fig.~\ref{Geometry Top view}(a) for one possible configuration).  One potential applied scenario of this configuration would be a strategy to establish and maintain a communication behind an obstacle located in $D_1$. In our simulation, we set the wavenumber to $k=10$ and consider $D_1\Subset W_1$ where $W_1$ is an annular sector given in the spherical coordinates (with respect to the origin) by
\begin{equation}
\begin{split}
        W_1  = & \left \{  (r, \theta, \phi): r \in [0.4, 0.7] , \theta \in \left[ \frac{\pi}{4} ,            \frac{3\pi}{4} \right] ,  \right. \left. \phi \in  \left[\frac{3\pi}{4} , \frac{5\pi}{4} \right] \right\}.
\end{split}
\label{W1}
\end{equation}
The far-field direction is exactly behind the near control, i.e., $\Bx_1 = (r, \theta, \phi) = (r,\frac{\pi}{2}, \pi)$, for large $r$. The desired field in region $D_1$ is $f_1 = 0$ while the desired far field pattern in direction $\Bx_1$ is given by $f_{\infty , 1} = 0.01 + i \cdot 0.02$ and $i = \sqrt{-1}$. The simulation results are shown in Fig.~\ref{Free Space Near Control}, Fig.~\ref{Free Space Far Control} and Fig.~\ref{Surface Normal velocity}.

The pointwise magnitude of the generated field in the near control is shown in Fig.~\ref{Free Space Near Control}. As a numerical stability check, this field is computed and plotted using points slightly off from the mesh points used in the collocation scheme. The generated field in the near control region is approximately a null with maximum absolute values less than $1.6 \times 10^{-8}$ and $L ^2$ norm of the generated field equal to $6.6941 \times 10^{-8}$.
Fig.~\ref{Free Space Far Control} shows the simulation results in the far-field directions. We consider a small patch around the exact direction $\Bx_1$. The two plots in Fig.~\ref{Free Space Far Control} show the pointwise relative errors of the real and imaginary parts. Note that the pointwise measure of error $e_{i}$ is defined as
\begin{equation}
\label{Pointwise error definition}
e_i =
\begin{cases}
     \frac{| u_i - f_i|} {| f_i|} &\  \text{ if } f_i \ne 0, \\
    |u_i - f_i| &\text{ if } f_i = 0,
\end{cases}
 \end{equation}
where $u = A w_d$ is the generated field and $f_i$ is the prescribed value in the $i^{th}$ evaluation point. Here, we find that the maximum pointwise error of the generated real and imaginary parts are both within order $10^{-6}$. In the exact direction $\Bx_1$, the relative errors of the real part and imaginary part are $4.7907 \times 10^{-8}$ and $1.6783 \times 10^{-8}$, suggesting good control accuracy.

Fig.~\ref{Surface Normal velocity} shows the normal velocity $v_n$ on the source $\partial D_a$ in the rectangular $(\theta, \phi)$-plot. The pointwise amplitudes of $v_n$ are quite small and within order $10^{-2}$. The average power and stored energy in the actual source are 8.2197 dB and 9.1525 dB, respectively, which implies that it is feasible for physical implementation.
%=========================================================================================
\subsection{\label{subsec:3:2} Two near control and two far field patterns}
\begin{figure}[hb]
\centering
    \subfigure[]{
        \includegraphics[width=0.45\linewidth]{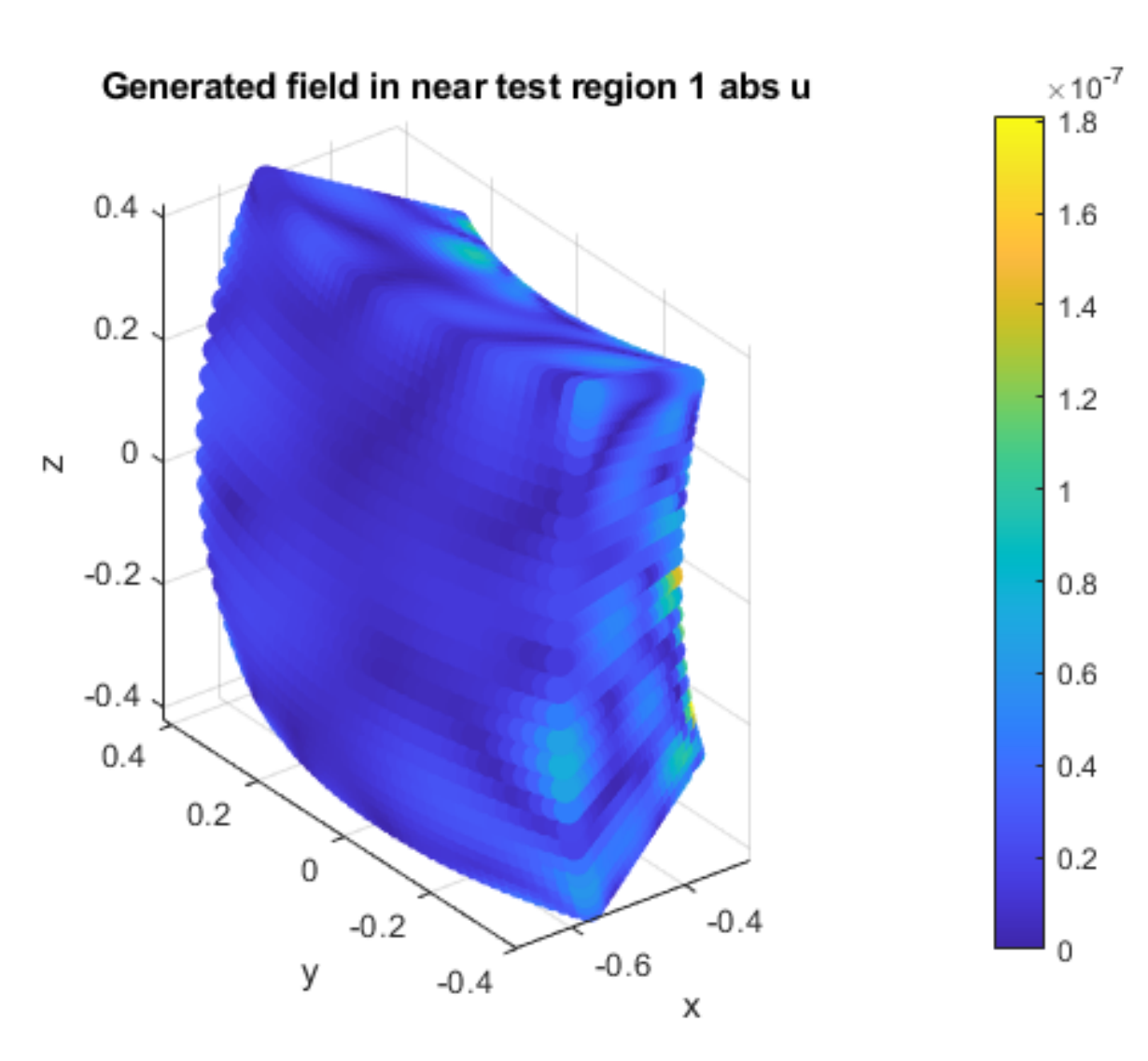}}
  \subfigure[]{
        \includegraphics[width=0.45\linewidth]{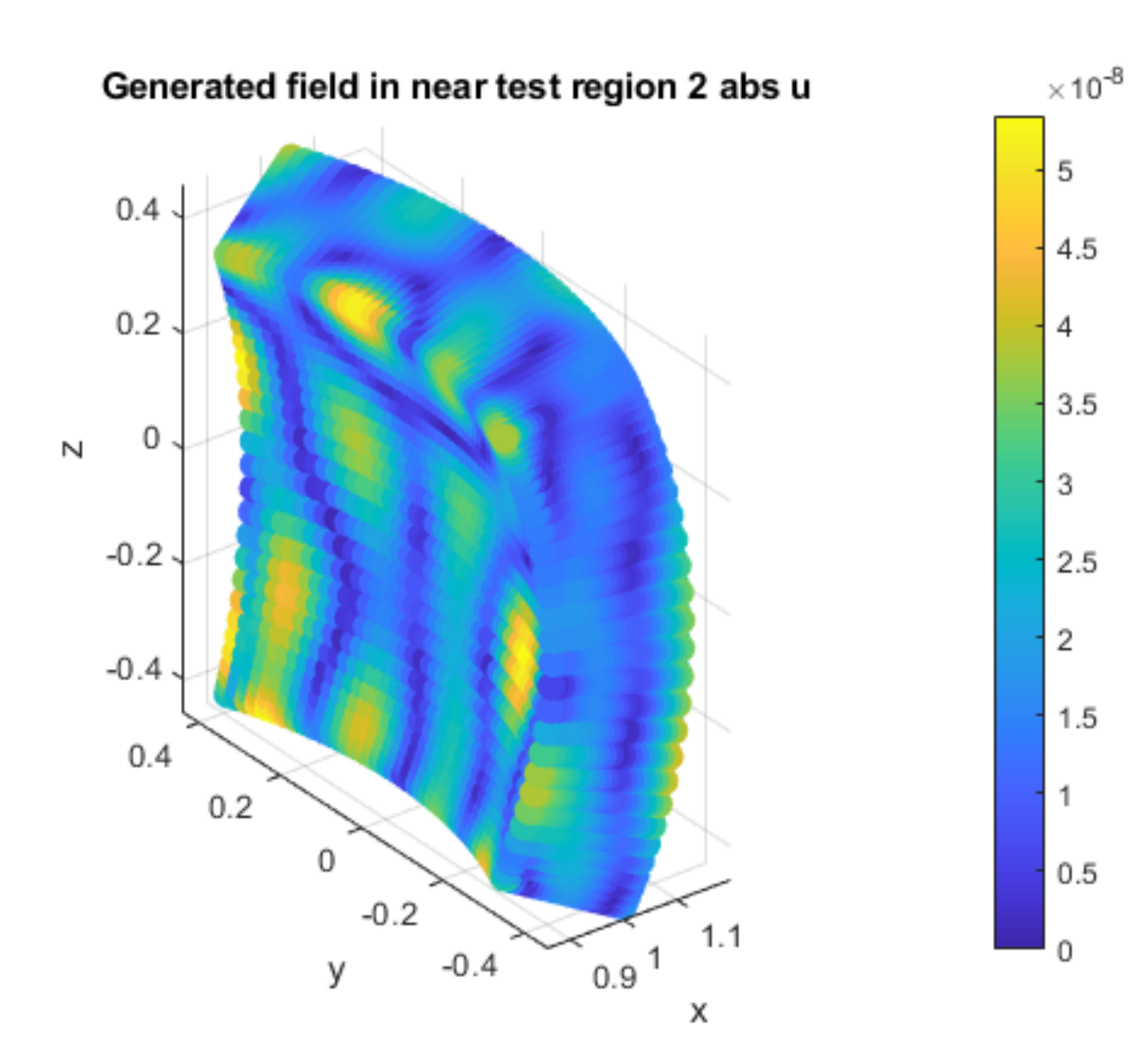}}
\caption{Pointwise magnitudes of the generated fields on (a) $D_1$ and (b) $D_2$, both approximating a null field.}
\label{Free Space Two Near Generated Field}
\end{figure}

\begin{figure}[!t]
\centering
\includegraphics[width=0.75\linewidth]{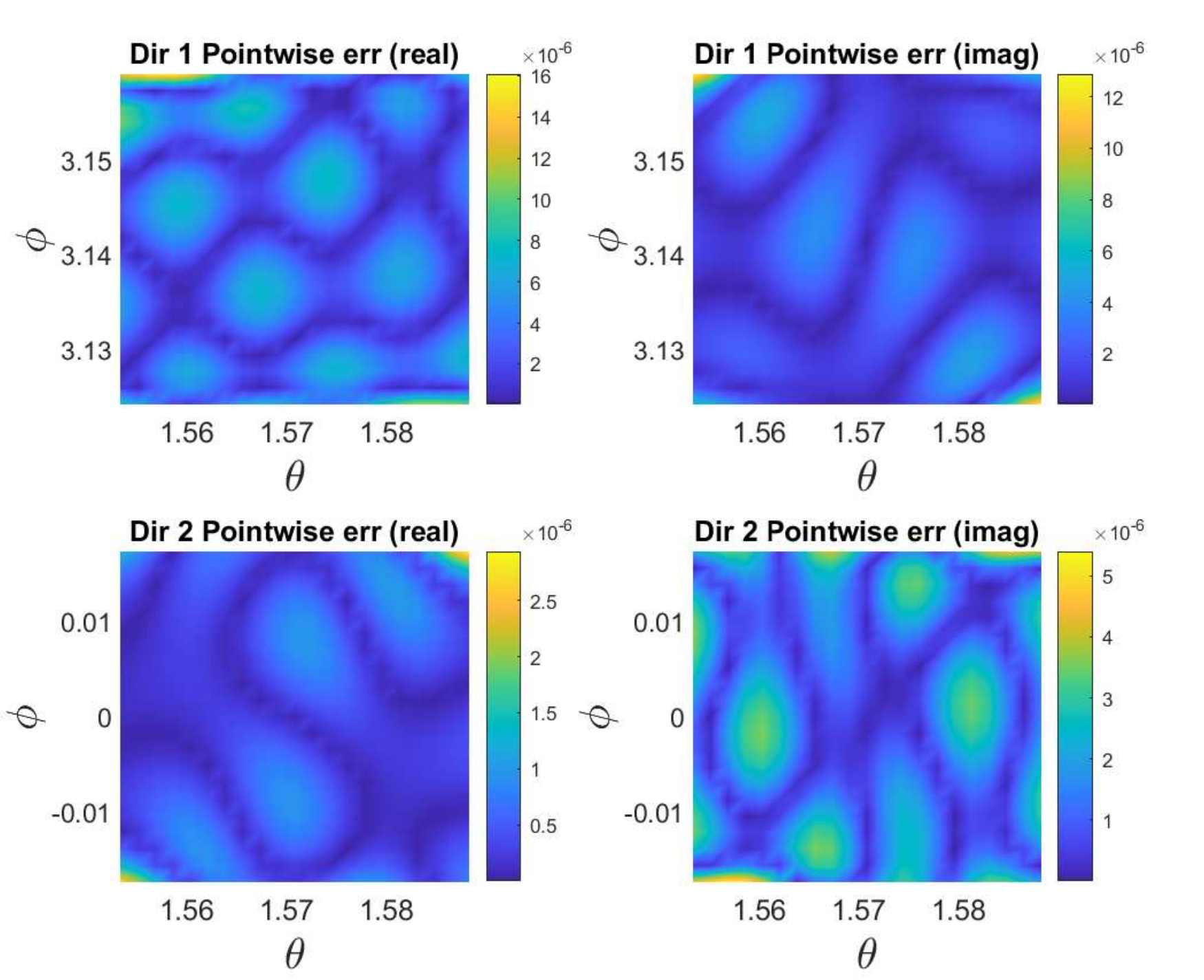}
\caption{Pointwise relative errors (real and imaginary parts) in patches around the two far-field directions.}
\label{Free Space Far Control Two Near}
\end{figure}

\begin{figure}[!hb]
\centering
    \includegraphics[width=0.5\linewidth]{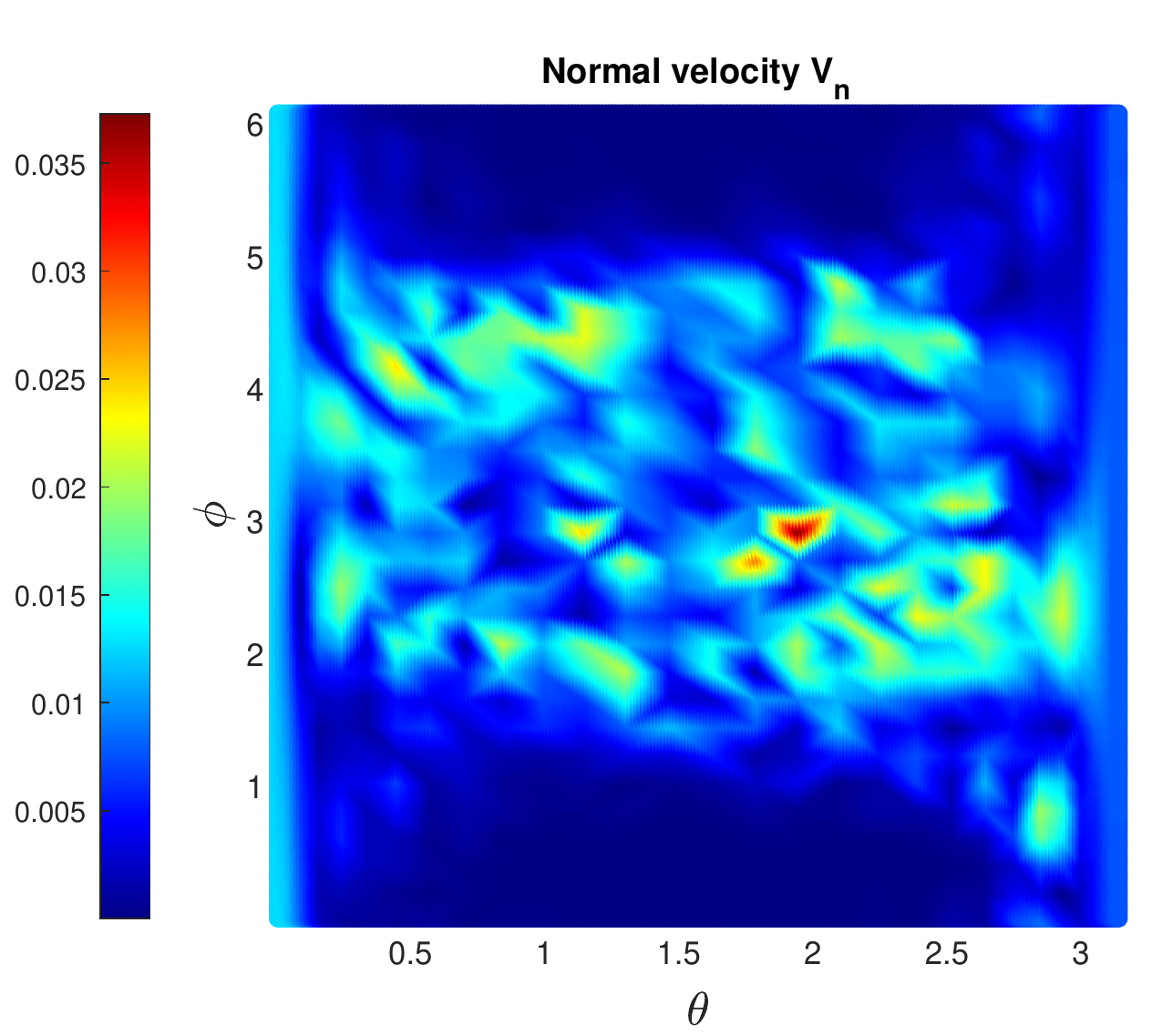}
\caption{Normal velocity $v_n$ on the actual source $\partial D_a$.}
\label{Free Space Surface Density and Normal velocity Two Near}
\end{figure}

In this subsection we show the performance of our scheme in creating null fields in $D_1$ and $D_2$ while approximating two distinct prescribed patterns $f_{\infty , 1}, f_{\infty , 2}$ in far-field directions $\Bx_1$ and $\Bx_2$, respectively (see Fig.~\ref{Geometry Top view}(b) for one possible configuration). One possible application of this configuration would be a strategy to establish and maintain communication in several given far-field directions while avoiding near field interference located in $D_1$ and $D_2$. In this test, $k=10$, $D_1\Subset W_1$ with $W_1$ defined in~\eqref{W1} while $D_2\Subset W_2$ with $W_2$ given in the spherical coordinates as
\begin{equation}
\begin{split}
        W_2  = & \left \{  (r, \theta, \phi): r \in [1.0, 1.2] , \theta \in \left[ \frac{\pi}{4} ,            \frac{3\pi}{4} \right] ,  \right. \left. \phi \in  \left[-\frac{\pi}{4} , \frac{\pi}{4} \right] \right\}.
\end{split}
\label{W2}
\end{equation}
In this geometry, $\Bx_1 = (r, \theta, \phi)= (r , \frac{\pi}{2}, \pi)$ and $\Bx_2= (r, \theta, \phi) = ( r,\frac{\pi}{2}, 0) $, $r \gg 1$ while the prescribed far-field pattern values are, $f_{\infty , 1} = 0.01 + i \cdot 0.02$ in direction $\Bx_1$ and $f_{\infty , 2} = 0.05 + i \cdot 0.03$ in direction $\Bx_2$.  The simulation results are shown in Fig.~\ref{Free Space Two Near Generated Field}, Fig.~\ref{Free Space Far Control Two Near} and Fig.~\ref{Free Space Surface Density and Normal velocity Two Near}. In Fig.~\ref{Free Space Two Near Generated Field}, we see that the absolute errors in the two near controls are within order $10^{-7}$. In Fig.~\ref{Free Space Far Control Two Near} we present the good performance in  two small patches of the far field centered in each of the two prescribed far field directions. In fact, when measured exactly in $\Bx_1$ and $\Bx_2$, the relative errors of the real parts are $3.2925 \times 10^{-7}$ and $6.3876 \times 10^{-8}$ while the relative errors of the imaginary part are $8.0229 \times 10^{-8}$ and $1.6111 \times 10^{-7}$, respectively. In Fig.~\ref{Free Space Surface Density and Normal velocity Two Near} we present the normal velocity required on the actual physical source $D_a$. The average radiated power is 8.3986 dB and the stored energy is 9.6504 dB. The power budget is slightly larger than that in Section~\ref{subsec:3:1} as the source is projecting two non-zero far-field patterns in two directions.
%=========================================================================================
\subsection{\label{subsec:3:3} Sensitivity analysis}
\begin{figure}[!b]
\centering
    \subfigure[]{
        \includegraphics[width=0.35\linewidth]{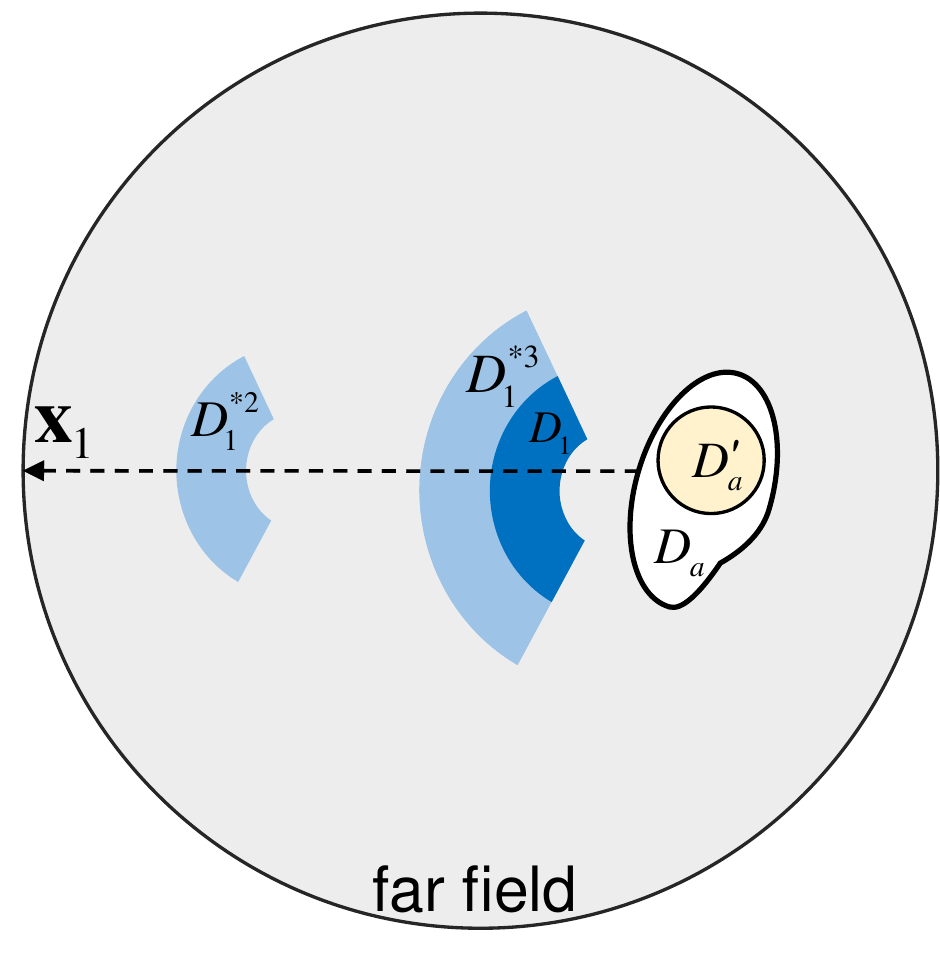}}
  \subfigure[]{
        \includegraphics[width=0.35\linewidth]{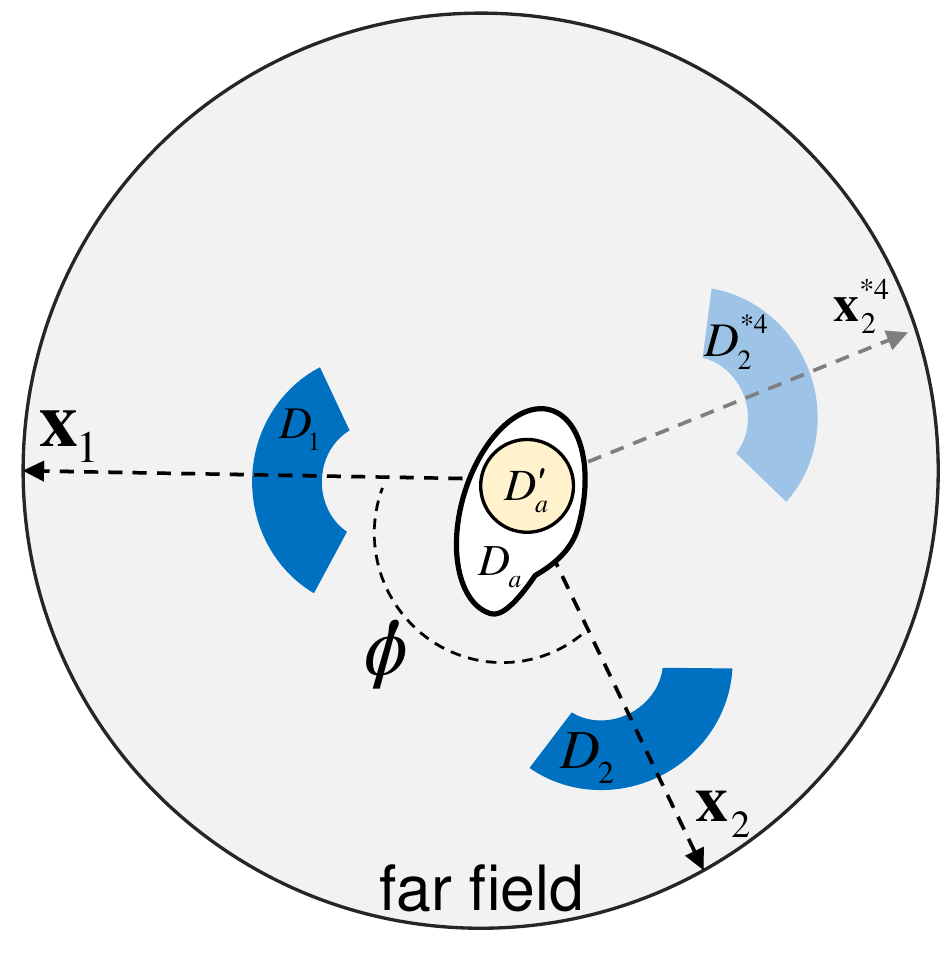}}
\caption{Sketch  of the geometry in sensitivity analysis. $D_1$, $D_2$ and $\Bx_1$, $\Bx_2$ are original near controls and far field directions, respectively. They are shown in dark color. The light-colored regions with $D^{*n}_1$, where $n = 2, 3$ and $4$, are corresponding to the experiments in Section~\ref{subsec:3:3:2} to Section~\ref{subsec:3:3:4}.}
\label{Sweep Geometry}
\end{figure}

The aforementioned results support the analysis of ~\cite{egarguin2020active} and show that our strategy produces good results for each of the two configurations depicted in Fig.~\ref{Geometry Top view}. In the following tests, we aim to study the sensitivity of our strategy with respect to variations in several physically relevant parameters, such as wavenumber $k$, the distance between the control region and the active source, the control region size and, mutual distance between the control regions (in the case of more control regions and far field directions Fig.~\ref{Geometry Top view}(b)). The feasibility of the active control scheme is also discussed by looking at the overall control accuracy, power budget, and anti-noise performance. The geometry in the sensitivity analysis is depicted in Fig.~\ref{Sweep Geometry}. To distinguish the original region (dark), the regions with modified parameters are shown in a light color. For instance, $D^{*2}_1$ denotes the near control which is shifted away from the active source, in which the superscript `2' is corresponds to the experiment number in Section~\ref{subsec:3:3:2}.
\subsubsection{\label{subsec:3:3:1} Varying the wavenumber $k$}
\begin{figure*}[!t]
\baselineskip=12pt
\centering
    \includegraphics[width=1\linewidth]{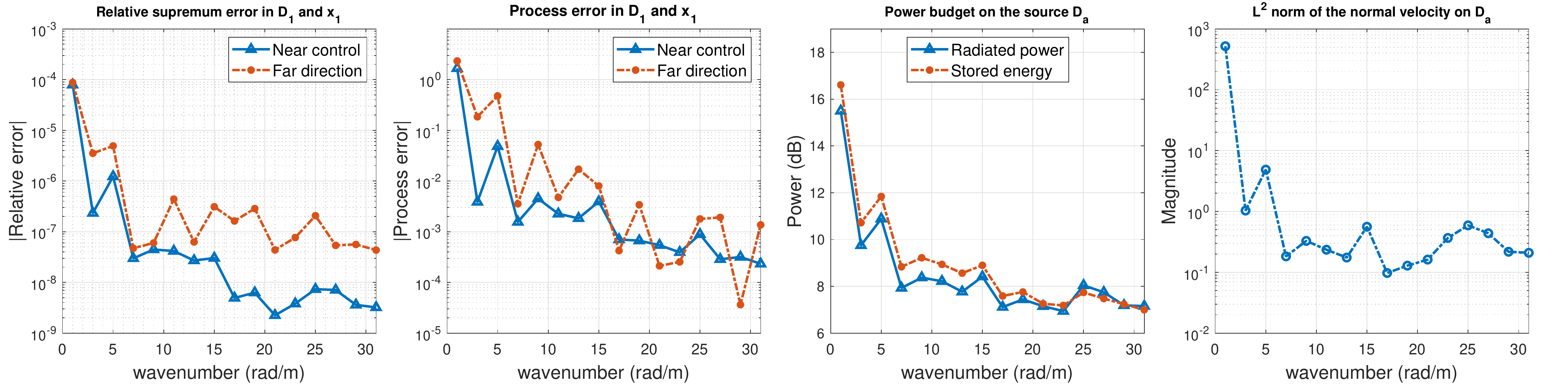}
\caption{Results showing the control accuracy and the power budget varying with $k$. From left to right, (1) Supremum error. (2) Process error. (3) Power budget on $D_a$. (4) $L^2$ norm of normal velocity $v_n$ on $D_a$.}
\label{Sweep k Free space One Null Near}
\end{figure*}

We start with the initial geometry in Fig.~\ref{Geometry Top view}(a), i.e., only one near control and a far-field direction. The prescribed field in $D_1$ is zero and the far-field pattern is a given non-zero complex number. In this simulation we let the wavenumber vary from 1 to 31. In every single simulation, we keep the geometry fixed and only change the wavenumber. The simulation results are shown in Fig.~\ref{Sweep k Free space One Null Near}. The first plot in Fig.~\ref{Sweep k Free space One Null Near} shows the supremum pointwise absolute error in $D_1$ and the relative error in the far field direction $\Bx_1$. The results indicate that these errors decrease as the wavenumber (frequency) increases. The second plot is the process error, which is the supremum pointwise  error when the manufacturing noise at the source or feeding error is considered. Mathematically, the process error can be obtained by replacing $u$ with $u_{\delta}$ in~(\ref{Pointwise error definition}), where $u_\delta = A w_{d\delta} $ and $\delta = 0.0001$ is the noise threshold. In this paper, we consider a random Gaussian noise such that $w_{d \delta} = w_d \cdot  (1 + \delta \cdot {\Vert w_d\Vert_{2}} \cdot \cal R )$ and $\cal R \sim \mathcal{N}$(0,1). The process error in Fig.~\ref{Sweep k Free space One Null Near} indicates that the system is capable of overcoming some noise at a certain level, either from manufacturing error or feeding network noise. The power budget and $L^2$ norm of $v_n$ show that the control effort decreases as the operating frequency increases.
% \begin{equation}
%     \text{Process error} = \sup_{u_\delta,f \in D_1} \left\{ \text{P.W. error} \right\}
% \label{Process error definition}
% \end{equation}
\subsubsection{\label{subsec:3:3:2} Varying the distance between the near control and the active source}
\begin{figure*}[ht]
\baselineskip=12pt
\centering
    \includegraphics[width=1\linewidth]{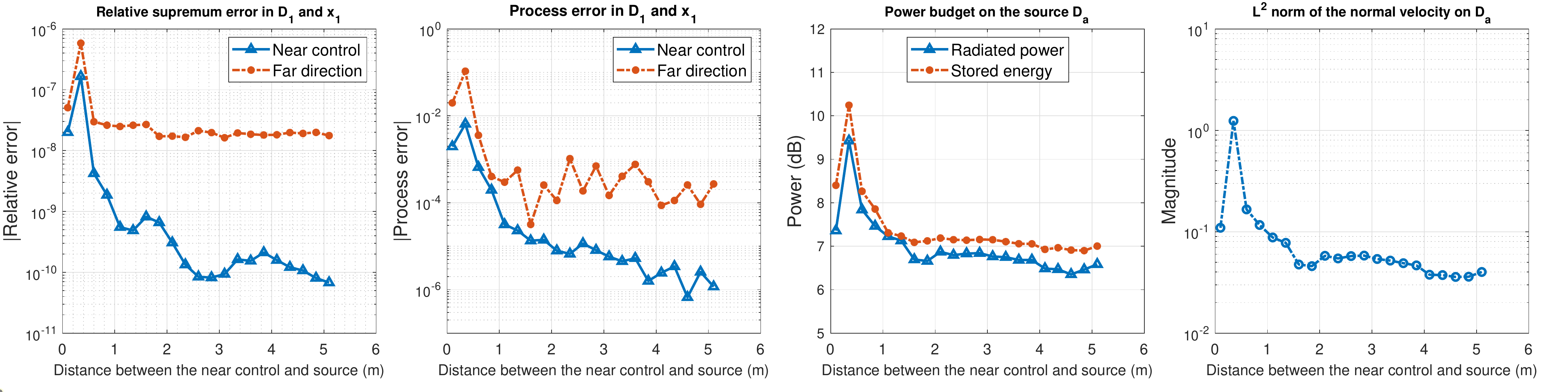}
\caption{Results showing the control accuracy and the power budget varying with mutual distance between $D_1$ and $D'_a$. From left to right, (1) Supremum error. (2) Process error. (3) Power budget on $D_a$. (4) $L^2$ norm of normal velocity $v_n$ on $D_a$.}
\label{Sweep distance Free space One Null Near}
\end{figure*}

The following results show the effect of variation in mutual distance between the near control and the active source. We still use the initial model in Fig.~\ref{Geometry Top view}(a). The near control $D_1$ is shifted further away from the source ($D^{*2}_1$ in Fig.~\ref{Sweep Geometry}(a)) and the other parameters are fixed. The results are depicted in Fig.~\ref{Sweep distance Free space One Null Near}. We notice that the control accuracy in the near region keeps on improving when the near control is moved further away from the source, while the control accuracy in the far direction is converging to a certain value. This indicates that the effect of the obstacle is weaker as $D_1$ is further away.  The power budget on the active source consequently follows a similar trend as the control accuracy.
\subsubsection{\label{subsec:3:3:3} Varying the near control size}
\begin{figure*}[ht]
\baselineskip=12pt
\centering
    \includegraphics[width=1\linewidth]{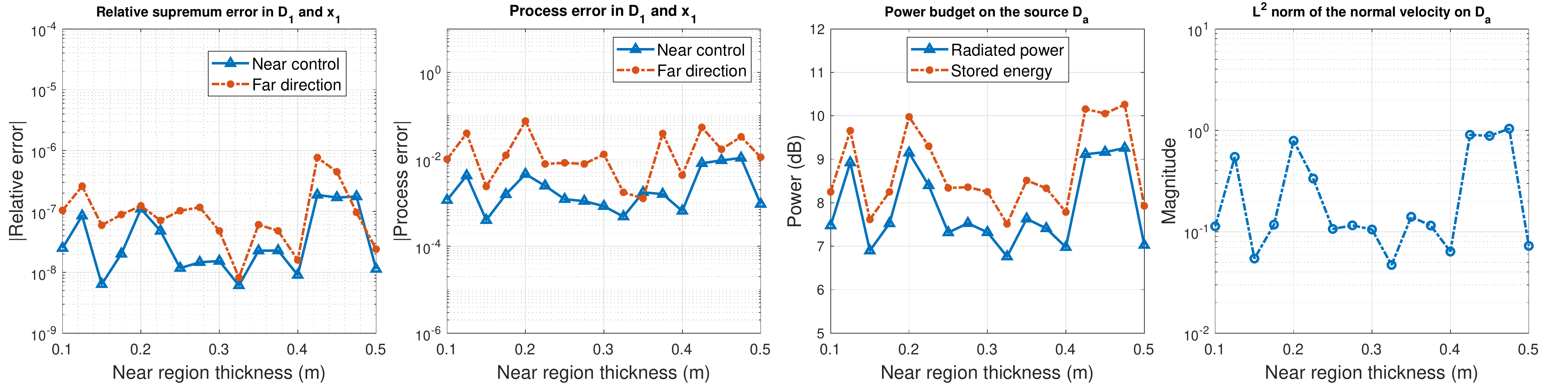}
\caption{Results showing the control accuracy and the power budget varying with the size of the near control $D_1$. The size of $D_1$ is the difference between the outer and inner radii, i.e., the thickness of the sectorial region. From left to right, (1) Supremum error. (2) Process error. (3) Power budget on $D_a$. (4) $L^2$ norm of normal velocity $v_n$ on $D_a$.}
\label{Sweep thickness Free space One Null Near}
\end{figure*}

Now we consider the behavior of the control accuracy and the power budget with respect to incremental increase in the outer radius of the near control region $D_1$ ($D^{*3}_1$ in Fig.~\ref{Sweep Geometry}(a)) with all the other parameters kept fixed. The results are shown in Fig.~\ref{Sweep thickness Free space One Null Near}. Notice that the control accuracy and power budget are slightly oscillating in the entire range of near control size. The results indicate a good performance for larger obstacles.
\subsubsection{\label{subsec:3:3:4} Varying the mutual distance between near controls}
\begin{figure*}[ht]
\baselineskip=12pt
\centering
    \includegraphics[width=1\linewidth]{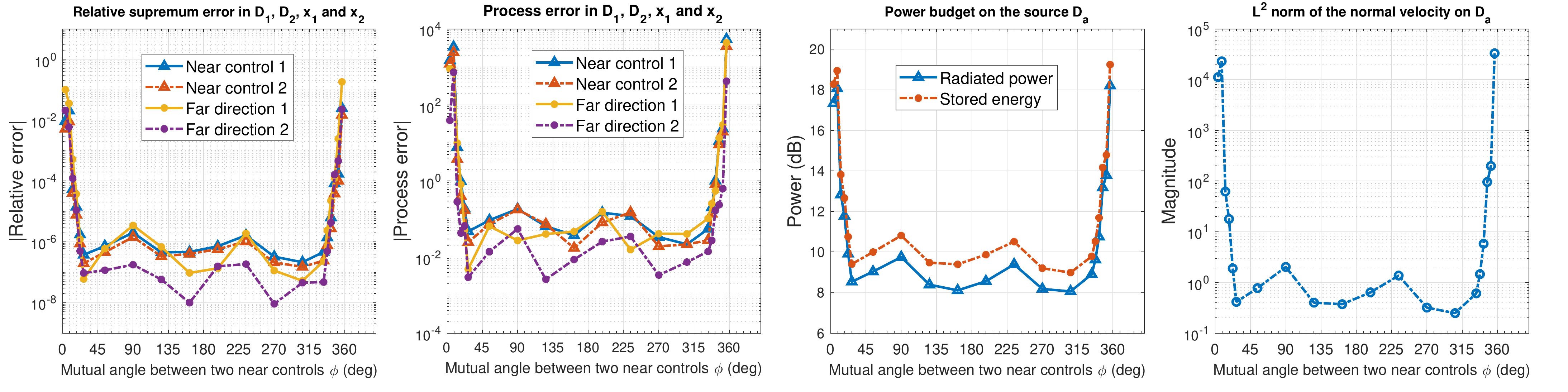}
\caption{Results showing the control accuracy and the power budget varying with the mutual distance between near controls $D_1$ and $D_2$ (given as a function of then angle $\phi$ between them). From left to right, (1) Supremum error. (2) Process error. (3) Power budget on $D_a$. (4) $L^2$ norm of normal velocity $v_n$ on $D_a$.}
\label{Sweep Angle Free space One Null Near}
\end{figure*}

In this sensitivity test, we consider two near control regions and vary the mutual distance between them. In Fig.~\ref{Sweep Geometry}(b), $D_1$ is fixed and we rotate $D_2$ around $D_a$ to obtain a new secondary control region $D^{*4}_2$. For a fair comparison, the far-field directions are kept exactly behind the controls, i.e., the second far field direction is also rotated (see $\Bx^{*4}_2$). The mutual distance between the two near controls is determined by the mutual angle $\phi$ shown in Fig.~\ref{Sweep Geometry}(b). The control accuracy and power budget are investigated by varying the angle $\phi$ from $3.6 ^{\circ}$ to $356.4 ^{\circ}$.The results are shown in Fig.~\ref{Sweep Angle Free space One Null Near}. We find that the control accuracy and power budget are within desired levels if $\phi$ is in the range of $[30^{\circ} , 330 ^{\circ}]$. If $\phi$ is out of this range, i.e., the near controls are too close to each other, the relative error and antenna power are blowing up. In the lower and upper bounds of the $\phi$-range, the relative error in the far-field patterns can be as high as 15\%. This indicates that the control regions cannot be too close, otherwise the accurate control effects are not guaranteed.
%=============================================================================
\section{\label{sec:4} Numerical results in the homogeneous ocean}
In this section, we extend our sensitivity analysis to the homogeneous ocean regime. We follow a similar procedure as in the previous section that deals with the problem in free space. We only show part of the numerical results due to the page limitation. The rest of the numerical examples is available in the supplementary document~\cite{suppl2020}. In the entire section, the fictitious source $D_a'$ is the sphere of radius 0.2 m with center at $(0, 0, -50)$ while, for exemplification,  the physical source is chosen to be the concentric sphere of radius 0.22 m. The control region $D_1$ and the far-field direction $\mathbf x_1$ are given by the control region and far-field direction described in the free space simulations but this time shifted $50$ m downward (see Fig.~\ref{Shallow water sketch} for a sketch). The depth of the ocean environment is $|h|=100$ m, the speed of sound $c$ is assumed to be 1515 m/s and the density $\rho$ is 1020 kg/m$^3$.
%==========================================================================================
\subsection{\label{subsec:4:1} A null near control and non-zero far field pattern}
\begin{figure}[!b]
\centering
    \includegraphics[width=0.5\linewidth]{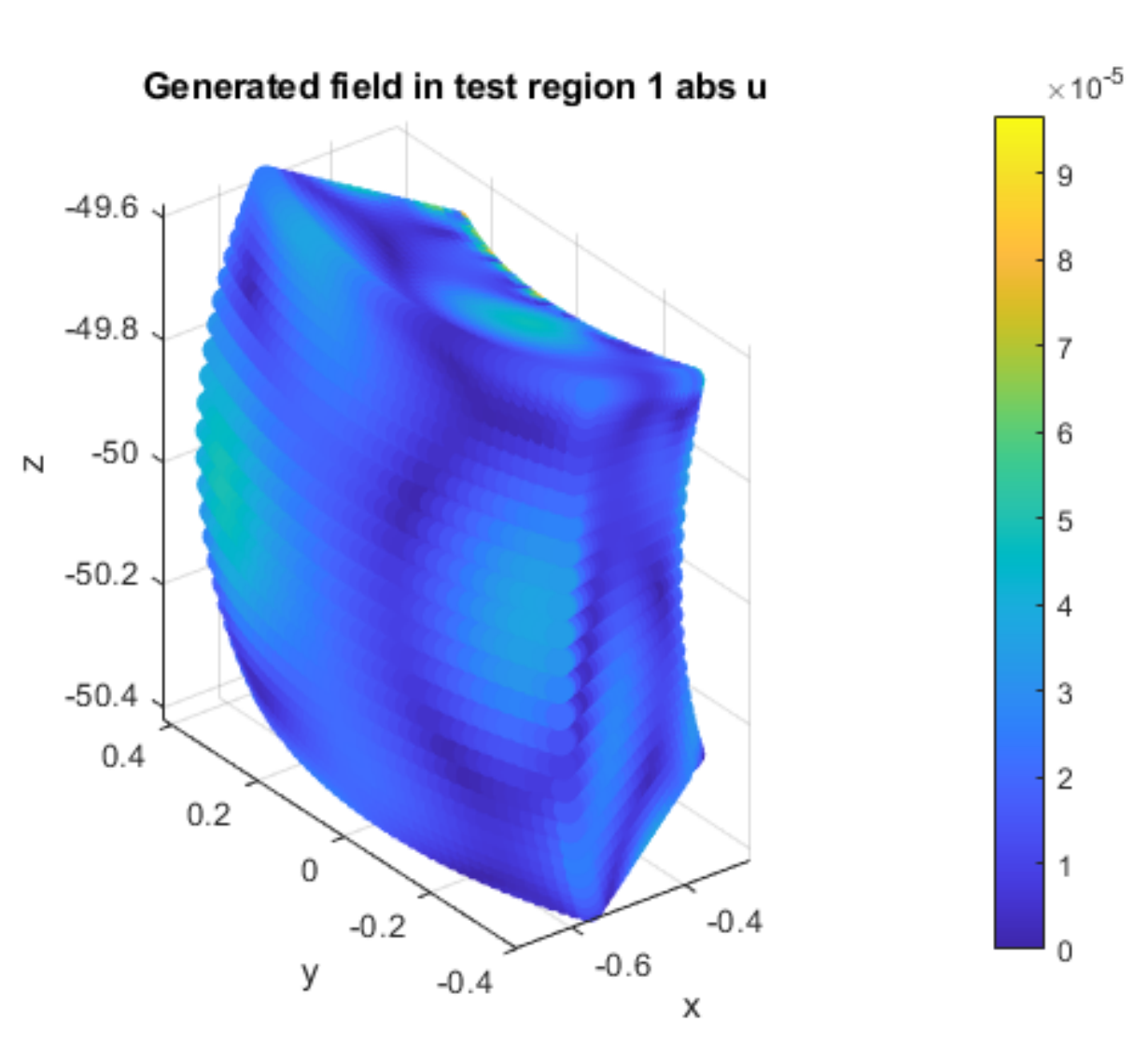}
\caption{Pointwise magnitude of the generated field in $D_1$ approximating a null field.}
\label{Homogeneous Ocean Near Control}
\end{figure}

\begin{figure}[!t]
\centering
    \subfigure[]{
        \includegraphics[width=0.4\linewidth]{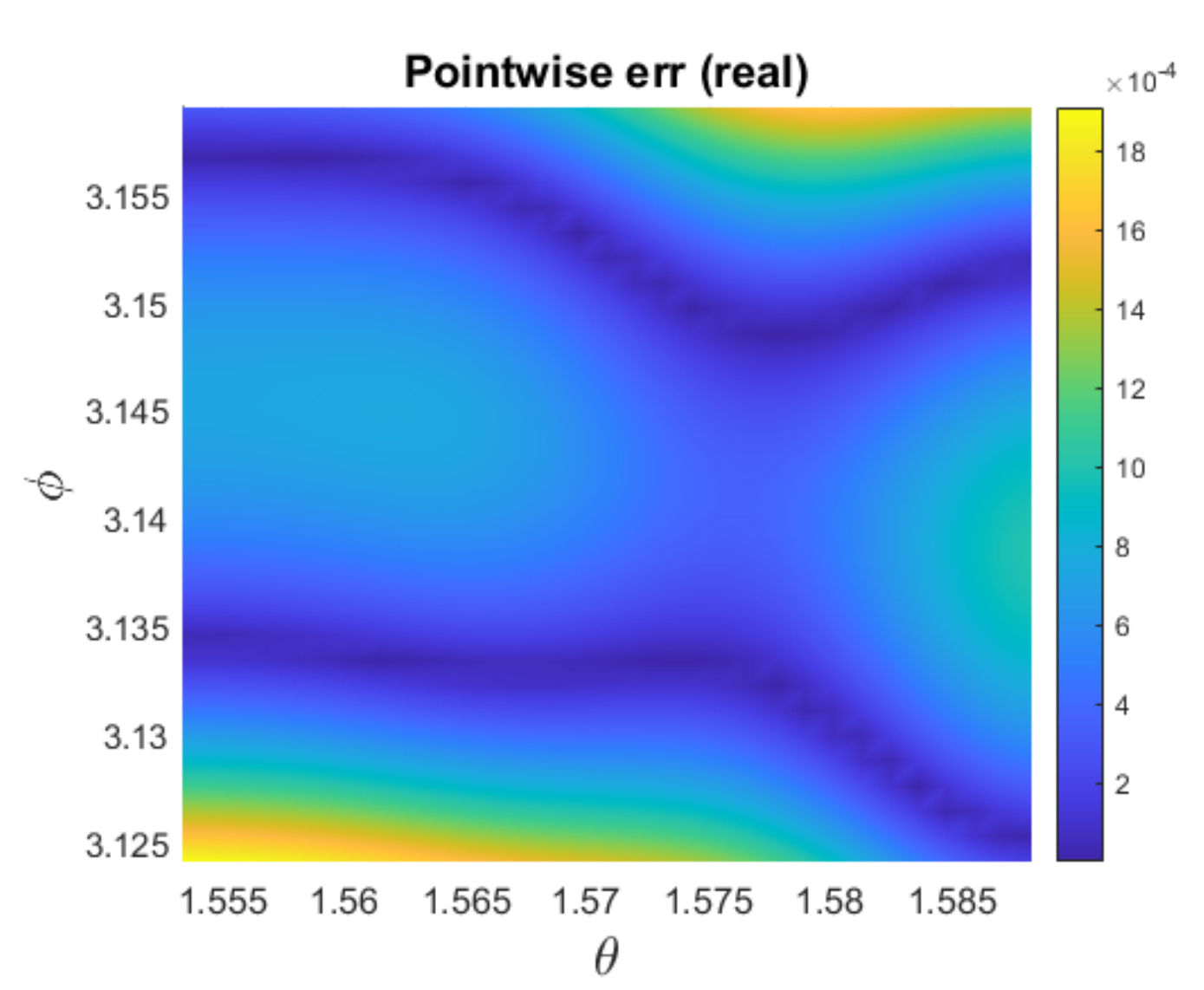}}
    \subfigure[]{
        \includegraphics[width=0.4\linewidth]{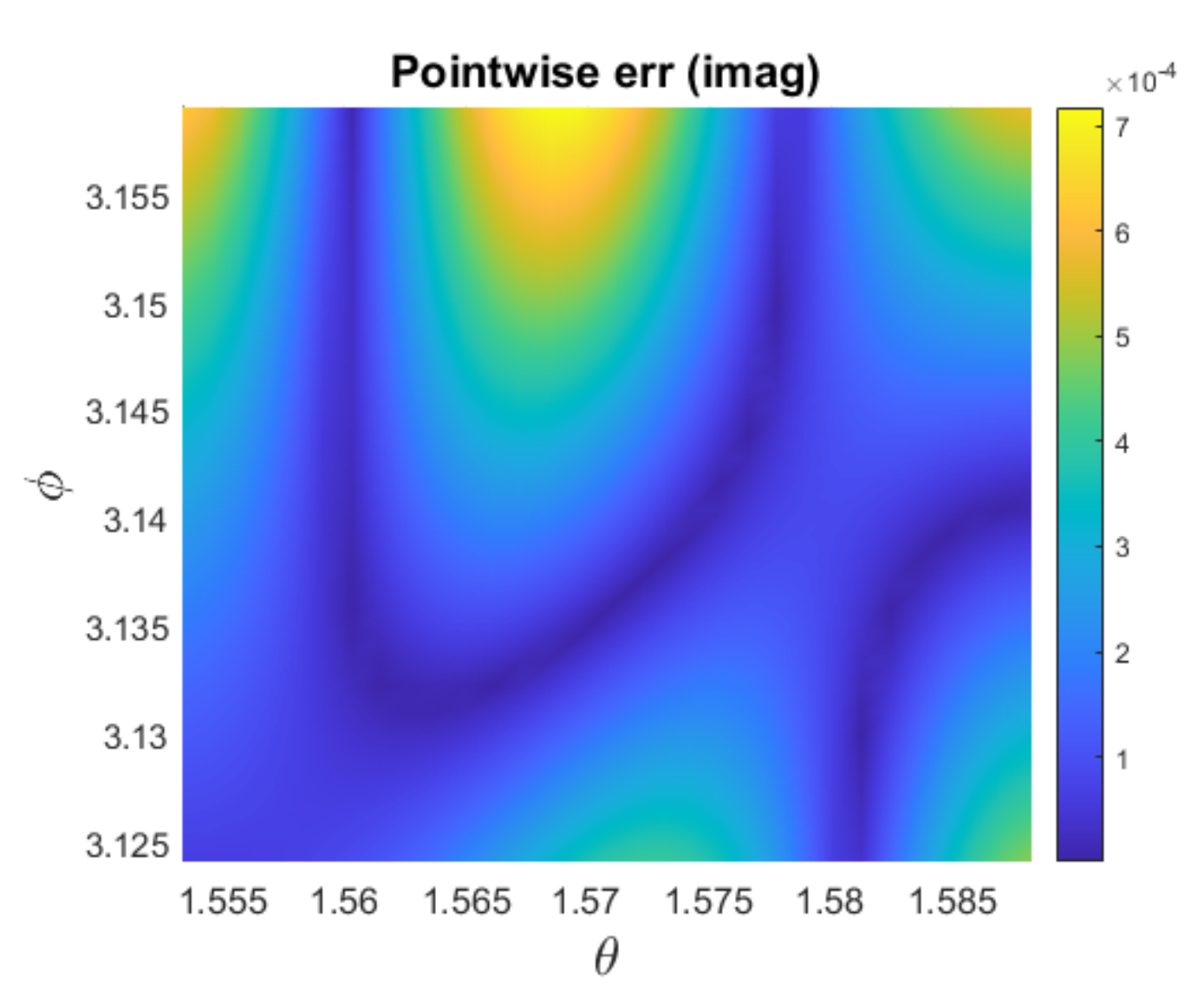}}
\caption{Pointwise relative error in a patch around the far field direction: (a) Real part, (b) Imaginary part.}
\label{Homogeneous Ocean Far Control}
\end{figure}

\begin{figure}[!t]
\centering
    \includegraphics[width=0.5\linewidth]{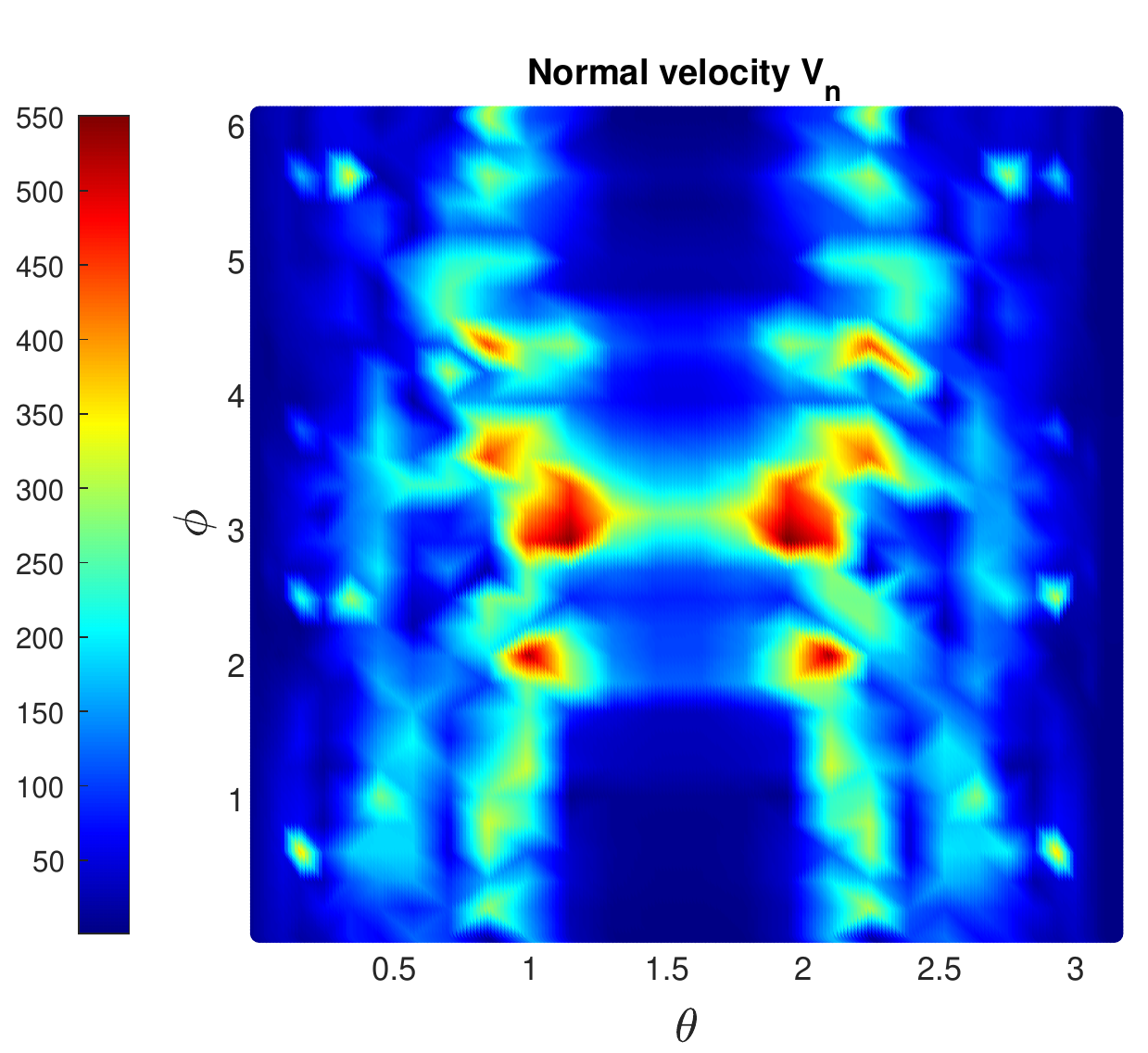}
\caption{Normal velocity $v_n$ on the actual source $\partial D_a$.}
\label{Homogeneous Ocean Surface Density and Normal velocity}
\end{figure}

We start from the initial geometry in Fig.~\ref{Geometry Top view}(a) but with $k=1$ and recall, in the spirit of~\cite{egarguin2020active}, the performance of our strategy. In $D_1$ we prescribe a null field, and the desired far-field pattern value is $f_{\infty , 1} = 0.01 + i \cdot 0.02$. The results are shown in Fig.~\ref{Homogeneous Ocean Near Control}, Fig.~\ref{Homogeneous Ocean Far Control} and Fig.~\ref{Homogeneous Ocean Surface Density and Normal velocity}. From Fig.~\ref{Homogeneous Ocean Near Control} we can observe that the maximum magnitude of the generated field in $D_1$ is within order $10^{-5}$. Similarly, Fig.~\ref{Homogeneous Ocean Far Control} shows the pointwise relative error in the real and imaginary parts of the far field pattern. In the exact direction $\Bx_1$, the relative errors of the real part and imaginary part are $5.2449 \times 10^{-4}$ and $1.8277 \times 10^{-4}$, respectively. In Fig.~\ref{Homogeneous Ocean Surface Density and Normal velocity}, we show the boundary input $v_n$ on the actual source $\partial D_a$. Consequently, the average power and stored energy on the actual source are 11.2158 dB and 11.5906 dB, respectively. The power budget at this level allows the physical source implementation.
\subsection{\label{subsec:4:2} Sensitivity analysis}
In this subsection we repeat the same experiments performed in the free space environment to explore the limitations or challenges of the active control in the homogeneous ocean environment. We only present here the effect of frequency and of the mutual distance between the near control and the source on the control accuracy and power budget. Some other experiment results are available in the supplementary material~\cite{suppl2020}.
\subsubsection{\label{subsec:4:2:1} Varying the wavenumber $k$}
\begin{figure*}[t]
\baselineskip=12pt
\centering
    \includegraphics[width=1\linewidth]{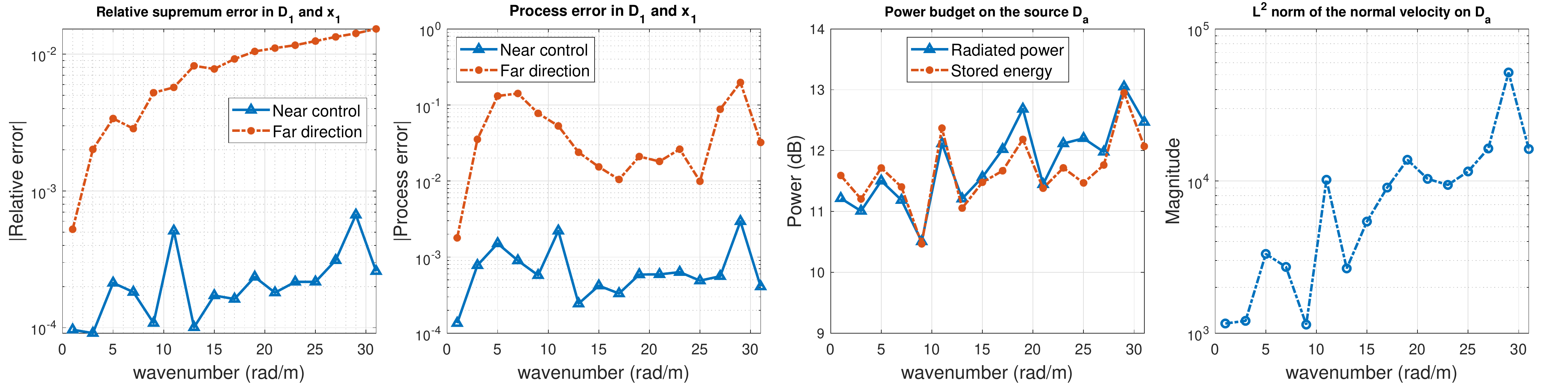}
\caption{Results showing the control accuracy and the power budget varying with wavenumber $k$. From left to right, (1) Supremum error. (2) Process error. (3) Power budget on $D_a$. (4) $L^2$ norm of normal velocity $v_n$ on $D_a$.}
\label{Sweep k Ocean One Null Near}
\end{figure*}

In this simulation, the wavenumber varies from 1 to 31 while the other parameters are fixed. The simulation results are shown in Fig.~\ref{Sweep k Ocean One Null Near}. In contrast to the free space results, the relative error increases as the frequency increases. This is due to the complex model (propagator, boundary condition, and so on) used in the homogeneous ocean environment. In practical underwater acoustic communication, the underwater channel poses serious challenges which are much more complicated than that in the free space. Hence, the control effort required on the source to accurately manipulate the near field and directional far field is more than that in the free space. Furthermore, the active scheme in the underwater environment is more sensitive to the manufacturing noise or feeding error. For the smaller noise threshold ($\delta = 10^{-8}$) the process error in the far direction is kept smaller than 10\%. This implies that even a low noise level in the feeding network may have a major effect on the far direction projection. However, the system still maintains a good performance in the near control. The overall trend of the power budget is also increasing as the wavenumber increases. It is noted that the spikes on the curves are due to the wavenumber being close to resonance frequencies.
\subsubsection{\label{subsec:4:2:2} Varying the distance between the near control and the active source}
\begin{figure*}[ht]
\centering
    \includegraphics[width=1\linewidth]{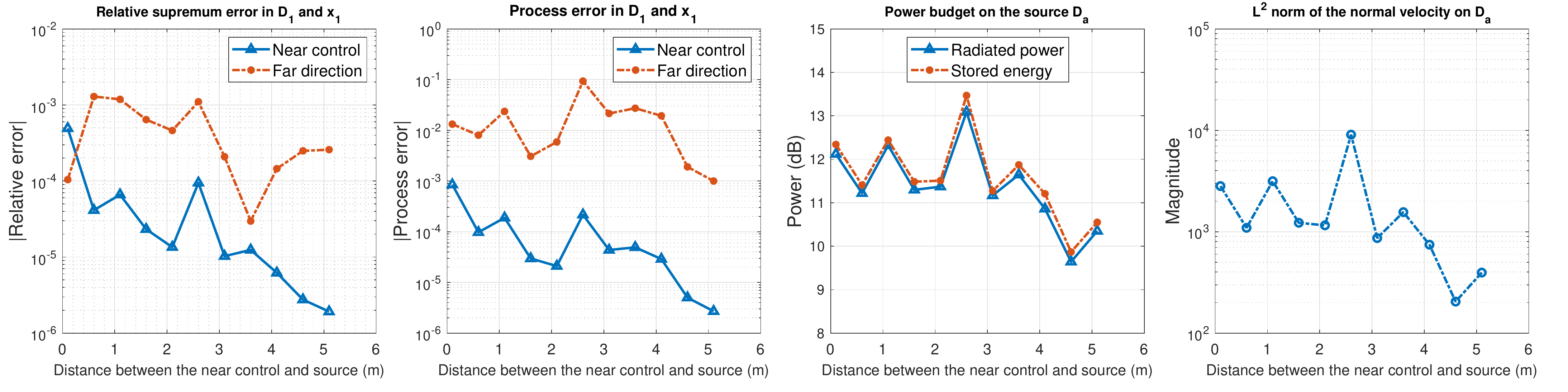}
\caption{Results showing the control accuracy and the power budget varying the distance between the source and the control region. From left to right, (1) Supremum error. (2) Process error. (3) Power budget on $D_a$. (4) $L^2$ norm of normal velocity $v_n$ on $D_a$.}
\label{Sweep distance Ocean One Null Near}
\end{figure*}

In this simulation, we test the sensitivity of our strategy as $D_1$ is moved away from the active source as shown in Fig.~\ref{Sweep Geometry}. The initial near control region is $D_1$ and it is incrementally pushed away from the source to obtain a new control region $D^{*2}_1$. The results are shown in Fig.~\ref{Sweep distance Ocean One Null Near}. We observe that the overall performance of control accuracy and power budget is better when the near control is further away from the active source, save some spikes due to resonances. The reason accounting for this trend is that the near control, acting as an obstacle is pushed away and hence the active source effort is eased.
%=============================================================================
\section{\label{sec:5}Conclusion}
In this paper, the feasibility of the active manipulation of Helmholtz fields both in free space and in a homogeneous ocean of constant depth is presented. We buld up on our previous works and demonstrated the possibility of feassible characterization of an active source (modeled as surface pressure or surface normal velocity) such that it is capable of approximating a priori given field in the near control while simultaneously projecting desired patterns in several far-field directions. We showed a good control accuracy and acceptable power budget of the proposed active control mechanism. Then, we explored the behavior of physically relevant parameters (power budget and control accuracy) with respect to variations in the frequency, outward shift, the outer radius of the near control and mutual distance between near controls.

In our simulations, we consider two initial models shown in Fig.~\ref{Geometry Top view}. The first one contains one near control and one far-field direction and the second one has two near control regions and two far-field directions. The far-field directions are places exactly behind the near control regions in each of the two models. In this paper, we only show the cases in which the near control(s) is(are) prescribed to be a null field(s) and we approximate a non-zero pattern(s) in the given far-field direction(s). This configuration mimic maintaining communication in desired far field directions while avoiding near field obstacles.

In free space, the control accuracy is within order $10^{-8}$ both in the near control and in the far field direction. Using the geometry shown in Fig.~\ref{Geometry Top view}(a), the operating frequency is first considered and the other parameters are fixed. The frequency is swept from 54.59 Hz to 1.69 kHz ($k$ is from 1 to 31.). The results indicate good performance even at the worst case considered. Moreover, the power requirement is kept at low levels which show the feasibility of a physical implementation for a practical source. In the second simulation, the near control region is moved outward. We notice that the control accuracy in the near region is continuously increasing as the near control region is pushed further away from the active source. However, accuracy converges to a certain level in the far-field direction. The results suggest that the active control scheme can handle the problem when the exterior region is either near or far from the source. Next, we vary the outer radius of the near control region to explore the effect of the near region's size on the control accuracy and power budget. The simulation results show that the active control scheme has a good performance in the entire range of the outer radii. We also consider the geometry in Fig.~\ref{Geometry Top view}(b). Here, we rotate the second near control together with the far-field direction behind it. We find that the control accuracy and power budget don't change significantly if the two near control regions are well-separated, in which the mutual angle $\phi \in [30^{\circ} , 300 ^{\circ}]$. However, outside this range the control performance is gradually degrading. The results suggest that realizing active control of two regions which are very close to each other is challenging.

Then, we extend our sensitivity analysis into the homogeneous constant depth ocean environment. The Green's function, corresponding to a pressure-release surface and a totally reflecting bottom, is expressed using the normal mode representation. Accordingly, the far-field pattern propagator is defined. In this case, the control accuracy is within orders $10^{-5}$ and $10^{-4}$ in the near control and in the far direction, respectively. In the first sensitivity analysis test we vary the wavenumber from 1 to 31. The results show that the overall performance of the control accuracy and the power budget decreases with increase in frequency. These suggest that the control scheme is more suitable to low frequencies in the homogeneous ocean environment. In the second test, we move the near control further away from the source. The results are similar to that in the free space, i.e., the control accuracy is higher and the power budget is lower as the near control is moved outward. More sensitivity studies are available in~\cite{suppl2020}.
%=============================================================================
\begin{acknowledgments}
This research has been supported in part by the Army Research Office under award number W911NF-17-1-0478 and in part by the National Science Foundation under award 1801925.
\end{acknowledgments}
%=============================================================================

%\bibliographystyle{unsrtnat}
\bibliographystyle{elsarticle-num}
\bibliography{ref}  %%% Remove comment to use the external .bib file (using bibtex).

\end{document}